\documentclass[aps,twocolumn,pra,tightenlines,floatfix,showpacs,superscriptaddress]{revtex4-1}

\usepackage{graphicx}
\usepackage{amsmath}
\usepackage{amssymb}
\usepackage{times}
\usepackage[english]{babel}
\usepackage{hyperref}

\begin{document}

\title{Critical temperature of trapped interacting bosons from large-$\mathcal{N}$ based theories}

\author{Tom Kim}
\affiliation{School of Natural Sciences, University of California, Merced, CA 95343, USA}

\author{Chih-Chun Chien}
\affiliation{School of Natural Sciences, University of California, Merced, CA 95343, USA}
\email{cchien5@ucmerced.edu}

\begin{abstract}
Ultracold atoms provide clues to an important many-body problem regarding the dependence of Bose-Einstein condensation (BEC) transition temperature $T_c$ on interactions. However, cold atoms are trapped in harmonic potentials and theoretical evaluations of the $T_c$ shift of trapped interacting Bose gases are challenging. While previous predictions of the leading-order shift have been confirmed, more recent experiments exhibit higher-order corrections beyond available mean-field theories. By implementing two large-$\mathcal{N}$ based theories with the local density approximation (LDA), we extract next-order corrections of the $T_c$ shift. The leading-order large-$\mathcal{N}$ theory produces results quantitatively different from the latest experimental data. The leading-order auxiliary field (LOAF) theory containing both normal and anomalous density fields captures the $T_c$ shift accurately in the weak interaction regime. However, the LOAF theory shows incompatible behavior with the LDA and forcing the LDA leads to density discontinuities in the trap profiles. We present a phenomenological model based on the LOAF theory, which repairs the incompatibility and provides a prediction of the $T_c$ shift in stronger interaction regime.
\end{abstract}


\maketitle
\section{Introduction}
Bose-Einstein condensation (BEC) occurs when low-energy states are macroscopically occupied by bosons below a critical temperature due to Bose-Einstein statistics, and the advent of ultracold atoms allows detailed analyses of BEC and its related phenomena \cite{pethick2008bose,ueda2010fundamentals,stoof2008ultracold,RevModPhys.76.599,RevModPhys.71.463,RevModPhys.80.885}. The BEC transition temperature, defined as the temperature when BEC starts to form, shifts with self-interactions of bosons and determining the functional form of the shift has been a challenge in many-body physics. The BEC transition temperature of a homogeneous Bose gas has been studied using various analytic and numerical methods (see Refs.~\cite{RevModPhys.76.599,RevModPhys.80.885} for a review). Experimental determinations of the dependence of the BEC temperature shift on self-interactions, however, have been complicated by the fact that ultracold atoms are usually trapped in optical or magnetic potentials. The trapping potential is usually of the harmonic form and leads to an inhomogeneous density profile of the atomic cloud. In noninteracting bosons, all particles fall to the ground state as the temperature approaches zero but this is no longer the case in interacting bosons since excitations out of the condensate can be finite even at zero temperature ~\cite{pethick2008bose,ueda2010fundamentals,stoof2008ultracold,RevModPhys.76.599,RevModPhys.71.463}.

In a harmonic trap, Bose-Einstein condensation starts to form at the trap center where the density is higher. On the other hand, repulsive interactions push particles away from each other and broaden the density profile. Hence, the density of repulsive bosons at the trap center is lower compared to a noninteracting system under similar conditions. Due to the reduction of the density at the trap center, the overall effect was proposed to be a negative $T_c$ shift as the interaction increases \cite{RevModPhys.71.463,BF02396737}. There have been experimental works on measuring $T_c$ of trapped interacting Bose gases. Ref.~\cite{PhysRevLett.92.030405} measured the $T_c$ shift by estimating the total atom number after integrating over a cloud image and deducing the condensate fraction using the Thomas-Fermi approximation. The critical temperature is inferred by using a time of flight method and the temperature is inferred by the size of the cloud after an expansion. In experiments, $T_c$ is considered as the temperature at which the deduced condensate fraction vanishes. The $T_c$ shift was estimated as $\delta T_c/T_c^0=\alpha N^{1/6}$, where $\alpha=-0.009$ with the two-body $s$-wave scattering length $a=5.31nm$ and the harmonic-trap length $a_H=1.00\mu m$. Ref.~\cite{PhysRevA.81.053632} prepared atomic clouds at different temperatures above as well as below $T_c$ and derived $T$ and $\mu$ from a fit to the Popov model~\cite{pethick2008bose}. It was reported that $\delta T_c/T_c^0=c^\prime \frac{a}{a_H}N^{1/6}$, where $c^\prime=-1.4$. Ref.~\cite{PhysRevLett.106.250403} prepared two atomic clouds concurrently, one with the targeted interaction and the other with a very small interaction, same trapping frequency, and very similar particle number as a reference point. The $T_c$ shift due to interactions was extracted by the difference of the results from the two clouds, which eliminates finite size effects. The result was $\delta T_c/T_c^0=b_1(a/\lambda_0)+b_2(a/\lambda_0)^2$ with $b_1=-3.5 \pm0.3$ and $b_2=46\pm5$. Here $\lambda_0$ is the thermal length of a noninteracting trapped Bose gas at its critical temperature.

On the theoretical side, by using the Popov approximation with the local density approximation (LDA)~\cite{PhysRevA.54.R4633}, the $T_c$ shift to the first order in interaction is found to be 
$\delta T_c/T_c^0=-3.426(a/\lambda_0)$,
which agrees well with later experimental results. 
To capture the higher-order corrections to the $T_c$ shift of trapped interacting bosons, we will implement a large-$\mathcal{N}$ expansion and its generalization called the leading-order auxiliary field (LOAF) theory, both of which are non-perturbative in interaction and temperature, along with the local density approximation (LDA). Here $\mathcal{N}$ counts the number of atomic species and the expansion already works reasonably when modeling ultracold bosons with $\mathcal{N}=1$ \cite{PhysRevLett.105.240402,ChienPRA12}.
 The large-$\mathcal{N}$ based theories, when applied to atomic Bose gases, are derivable from well-defined thermodynamic free energies, show a smooth second-order BEC transition, and are consistent with standard perturbation theory or renormalization analysis in low-temperature and weakly-interacting regimes~\cite{PhysRevLett.105.240402,PhysRevA.85.023631,ChienPRA12,ChienAnnPhys}. The normal state properties from the large-$\mathcal{N}$ expansion also compare favorable with Monte-Carlo simulations~\cite{PhysRevA.91.043631}. Here we will show that while the $T_c$ shift from the LOAF theory with the LDA captures the functional form from the most recent experimental data \cite{PhysRevLett.106.250403}, the LOAF theory exhibits behavior incompatible with the LDA. The deviation from the conventional LDA illustrates another challenge of applying mean-field theories to trapped interacting Bose gases, and we will discuss a phenomenological model based on the LOAF theory allowing us to extract the functional form of the $T_c$ shift in the stronger interaction regime.

The paper is organized as follows. We first check the validity of the LDA using trapped noninteracting Bose gases in Sec.~\ref{sec:LDA} and show that the exact $T_c$ as well as the first-order finite particle number correction can be reproduced in the LDA. Sec.~\ref{sec:largeN} summarizes the large-N expansion, the leading-order large-N theory, and the LOAF theory. Their integrations with the LDA are summarized in Sec.~\ref{sec:LargeNLDA}. The numerical results of the trapped profiles and $T_c$ shifts from the two theories are presented in Sec.~\ref{sec:Results} and compared to the latest experimental results. In the same section, incompatible behavior of the LOAF theory with the LDA is discussed and a phenomenological model based on the LOAF theory is presented, which provides further theoretical predictions. Finally, Sec.~\ref{sec:conclusion} concludes our work.

\section{Local density approximation}\label{sec:LDA}
To evaluate the $T_c$ shift of a trapped Bose gas, the full density profile needs to be constructed because the emergence of BEC at the trap center depends on the particle density at the center, which has to be determined from a consistent density profile in the whole trap. While most mean-field theories are designed for uniform quantum gases, a powerful tool called the local density approximation (LDA) allows one to construct a full density profile by slicing the system into pieces and treating each piece as a locally uniform system \cite{RevModPhys.71.463,pethick2008bose}. By sewing all the pieces with a smooth profile of the chemical potential, an approximated density profile can be obtained. The LDA has been applied to trapped quantum gases and proved to be a versatile treatment \cite{RevModPhys.71.463,stoof2008ultracold,RevModPhys.80.885,PhysRevLett.98.110404,Stewart2008}.

\subsection{$T_c$ of noninteracting bosons}
At first look the LDA may not accurately describe finite-temperature phenomena like the BEC transition. We begin by checking the validity of LDA for a noninteracting trapped Bose gas at its transition temperature. According to the Bose-Einstein statistics, the number of bosons in excited states is given by
\begin{equation}
N_T=\sum_{\epsilon} \frac{1}{\exp(\beta\left(\epsilon-\mu\right))-1}.
\label{eq:BCE}
\end{equation}
Here $\mu$ is the chemical potential, $\beta=1/(k_B T)$ and we set the Boltzmann constant $k_B=1$.
 For a homogeneous noninteracting Bose gas, the BEC temperature is the lowest $T$ satisfying $N_T/\Omega=\rho$ with $\mu=\epsilon_0$ in the thermodynamic limit, where $\rho$ is the particle density, $\Omega$ is the system volume, and $\epsilon_0$ is the single-particle ground-state energy. For a parabolic energy dispersion $\epsilon_k=\hbar^2 k^2/(2m)$ with wave vector $k$, the Planck constant divided by $2\pi$, $\hbar$, and particle mass $m$, it can be shown that \cite{fetter2012quantum}
\begin{equation}
T_c^{0,homo}=\left(\frac{\rho}{\zeta(3/2)}\right)^{2/3}\frac{2\pi \hbar^2}{ m k_B},
\end{equation}
where the superscript $0$ denotes quantities of a noninteracting Bose gas and $\zeta(x)$ is the Riemann zeta function.

For noninteracting bosons in a harmonic trap, the energy eigenvalues are  $E_{n_1,n_2,n_3}=\hbar\omega\left(n_1+n_2+n_3\right)+\frac{3}{2}\hbar\omega$, where $\omega$ denotes the trap frequency. The total particle number is ~\cite{Haugerud199718,Grossmann1995188}
\begin{equation}\label{eq:ExactNonint}
N=\sum_{n_1,n_1,n_3} \frac{1}{\exp\left[\beta\hbar\omega\left(n_1+n_2+n_3\right)+\beta\left(E_0-\mu\right)\right]-1}. 
\end{equation}
$T_c^0$ is the lowest temperature when the equation is satisfied with $\mu=E_0=(3/2)\hbar\omega$ in the thermodynamic limit, which can be calculated analytically~\cite{mazenko2000equilibrium,BF02396737}. Explicitly, 
\begin{equation}\label{eq:Tc0trap}
T_c^0=\left(\frac{N}{\zeta(3)}\right)^{1/3}\frac{\hbar\omega}{k_B}.
\end{equation}

One may also apply the LDA to obtain the $T_c$ of a trapped noninteracting Bose gas. By approximating the energy dispersion with a parabolic one $\epsilon=E_0+\frac{\hbar^2 k^2}{2m}+\frac{1}{2} m \omega^2 r^2$ in Eq.~\eqref{eq:BCE} and replacing the summation by integrals over space and momentum, one obtains the number of bosons in the thermal cloud as  
\begin{equation}
N_T=\left(\frac{k_b T}{\hbar\omega}\right)^3 g_{3/2}(e^{(\mu-E_0)/{k_B T}}),
\end{equation}
where $\operatorname{g}_s(z) = \sum_{k=1}^\infty {z^k \over k^s} $ is the polylogarithm. We will redefine $\mu$ as the chemical potential measured from the zero-point energy of the system, so $\mu-E_0 \rightarrow \mu$ from here on. By expanding the series around $\beta\mu=0$ with $N_T=N$, the leading order gives Eq.~\eqref{eq:Tc0trap} and next order reproduces the leading-order finite-size correction presented in Eq.~\eqref{eq:deltaTc0}. Therefore, the validity of LDA has been established for noninteracting Bose gases.

Higher-order correction from finite $N$ can be found from Eq.~\eqref{eq:ExactNonint} by  
applying the Euler-Maclaurin formula~\cite{Haugerud199718,Grossmann1995188} or turning the summand into a geometric series and then carrying out the summation by expanding around $\beta \hbar \omega=0$ resulting in polylogarithmic functions~\cite{PhysRevA.54.656,PhysRevA.54.4188,PhysRevA.58.1490}. Explicitly, 
\begin{equation}
\frac{\delta T_c^0}{T_c^0}=-\frac{\zeta(2)}{2(\zeta(3))^{2/3}}N^{-1/3}\approx -0.73 N^{-1/3}.
\label{eq:deltaTc0}
\end{equation}
The correction decays with $N$ and is present regardless of self-interactions. In experiments those finite-size corrections are discarded by taking the difference of the results from two similarly prepared systems at different interaction strengths~\cite{PhysRevLett.106.250403}.

Ref.~\cite{PhysRevA.83.023616} derived the next order correction to the $T_c^0$ shift due to finite $N$ by using the LDA with an expansion of $\mu/k_B T$. However, Ref.~\cite{PhysRevA.92.017601} points out an ambiguity in the definition of the critical temperature because of finite-size effects. Thus, the next order correction from finite particle number may not provide a better pointer to the critical regime than the first order term when compared to numerical results. Here we will focus on systems with well-defined thermodynamic limit and will not include corrections from finite particle number already present in the noninteracting system in our later discussions.

\subsection{Interacting bosons}
The shift of $T_c$ from its noninteracting value $T_c^0$ in the presence of interactions has been a great challenge. Even for a uniform Bose gas, it took a long time for results from various studies to converge~\cite{PhysRevLett.106.250403,RevModPhys.76.599,RevModPhys.80.885}. The issue has been settled more recently and the leading-order shift is now believed to have a form of
$\Delta T_c^{homo}/T_c^{0,homo}= c \rho ^{1/3}a$,
where $a$ is the $s$-wave two-body scattering length and $c$ is a positive constant. Different values of $c$ have been reported using various analytic or numerical methods~\cite{RevModPhys.76.599}.

In a trapped Bose gas, a repulsive interaction flattens the density profile and lowers the density at the trap center~\cite{RevModPhys.71.463,PhysRevLett.106.250403}. As a consequence, the leading order of the $T_c$ shift for a trapped interacting Bose gas is believed to have the form 
$\frac{\delta T_c}{T_c^0}=c^{\prime}\frac{a}{a_{H}}N^{1/6}$, where $c^{\prime}$ should be a negative number. Here $a_{H}=\sqrt{\frac{\hbar}{m\omega}}$ is the harmonic length.
An early theoretical analysis using the Popov approximation~\cite{PhysRevA.54.R4633} provided an estimation of $c^{\prime}$ by introducing an approximated dispersion 
$\epsilon=\frac{\hbar^2 k^2}{2m}+V(r)+2\lambda n_T(r)-\mu$   
into Eq.~\eqref{eq:BCE}, where $V(r)=\frac{1}{2}m\omega^2r^2$ is the harmonic trap potential, $\lambda$ is the coupling constant, and $n_T(r)$ is the thermal particle density at radius $r$. After expanding the expression to the first order in $\delta T_c$, $\lambda$, and $\mu$, the $T_c$ shift to the first order due to interaction is found to be 
$\frac{\delta T_c}{T_c^0}\approx-1.33\frac{a}{a_{H}}N^{1/6}$, which agrees well with later experiments~\cite{PhysRevLett.106.250403}

An estimation of the second-order $T_c$ shift has been shown in Ref.~\cite{PhysRevA.64.053609} by expanding the distribution function in powers of the fugacity around $\beta\mu=0$, and the following expression was obtained. 
\begin{equation}
\frac{\delta T_c}{T_c^0}=c_1\frac{a}{\lambda _0}+\left(c_2^{\prime}\ln \left(\frac{a}{\lambda _0}\right)+c_2^{\prime \prime}\right)\left(\frac{a}{\lambda _0}\right)^2,
\end{equation}
where $c_1 =-3.426$, $c_2^{\prime} =-45.86$, and $c_2^{\prime\prime} =-155.0$. However, corrections from the logarithmic term were not reported in later experiments \cite{PhysRevLett.106.250403}.

Instead of assuming small interaction strength and expanding around the noninteracting limit, we use a path integral formalism to formulate trapped interacting bosons and apply the large-$\mathcal{N}$ expansion to find $T_c$. 
To handle the background harmonic trap, we adjust the theory to fit the LDA framework. Obtaining a full density profile in a trap is often a difficult task if a theory can only apply to a small range of temperature close to $T=0$ or $T=T_c$. This is because in a trap the local temperature scale $T/T_c^0(r)$ spans a wide range. Here $T_c^0(r)$ is the critical temperature of a noninteracting Bose gas with the same local density. Previous works on weakly interacting bosons have encountered challenges. For instance, the Popov theory exhibits an artificial first order transition at $T_c$~\cite{PhysRevLett.105.240402,popov1991functional},  and a higher-order large-$\mathcal{N}$ expansion used in Ref.~\cite{0295-5075-49-2-150} mainly focused on a uniform system near its critical temperature. 

The leading-order large-$\mathcal{N}$ theory and its generalization both exhibit second order transition and is not temperature restrictive \cite{ChienPRA12,PhysRevLett.105.240402}. By using the LDA, we calculate the trap density profile and estimate $T_c$ for trapped interacting Bose gases. To compare atomic clouds with the same total particle number, we impose the following condition to fix the total particle number $N$.
\begin{equation}\label{eq:number}
N=\int d^3 x \rho(x).
\end{equation}

\section{large-$\mathcal{N}$ based theories}\label{sec:largeN}
\subsection{Leading-order large-$\mathcal{N}$ theory}
The partition function of a single component Bose gas can be cast in an imaginary-time path-integral formalism~\cite{fetter2012quantum,ChienPRA12}. Explicitly, 
\begin{equation}
Z(\mu ,\beta,j)=\int D \phi D \phi ^* e^{-S\left(\phi ,\phi ^*,\mu ,\beta \right)+\int[dx](j^*\phi+j\phi^*)}.
\label{eq:partition}
\end{equation}
Here $\beta=(k_B T)^{-1}$, $\mu$ is the chemical potential, $[dx]\equiv d\tau d^{3}x$, and $j$ (or $j^*$) is the source of $\phi^*$ (or $\phi$). The action with imaginary time $\tau$ is 
$S\left(\phi ,\phi ^*,\mu ,\beta \right)=\int[dx]\mathcal{L}\left(\phi ,\phi ^*,\mu \right)$. 
For a nonrelativistic dilute Bose gas with contact interactions, the effective Euclidean Lagrangian density is
\begin{eqnarray}
\mathcal{L}&&=\frac{1}{2} \hbar  \left(\phi ^* \frac{\partial \phi }{\partial \tau }-\phi  \frac{\partial \phi ^*}{\partial \tau }\right)
-\frac{1}{2}\left(\phi ^*\frac{\hbar ^2\nabla ^2}{2 m}\phi +\phi \frac{\hbar ^2\nabla ^2}{2 m}\phi ^*\right) -\nonumber\\
&&\mu  \phi ^* \phi + \frac{1}{2} \lambda  \left(\phi ^* \phi \right)^2.
\label{eq:Lagrangian}
\end{eqnarray}
Here $\lambda$ is the bare coupling constant. 
In what follows we set $\hbar=1$, $k_B=1$, and $2m=1$. 
By introducing 
\begin{eqnarray}
\Phi &=&\left(
\begin{array}{cc}
 \phi,  & \phi ^* \\
\end{array}
\right)^{\mathsf{T}},J=\left(
\begin{array}{cc}
 j, & j^*  \\
\end{array}
\right)^{\mathsf{T}}, \nonumber \\
\bar{G}_0^{-1}&=&
\left(
\begin{array}{cc}
 \frac{\partial }{\partial \tau } -\frac{\nabla^2}{2m} & 0 \\
 0 & -\frac{\partial }{\partial \tau }- \frac{\nabla^2}{2m} \\
\end{array}
\right),
\end{eqnarray}
 $S$ can be written as
\begin{equation}
S=\int [dx] \left(\frac{1}{2}\Phi ^{\dagger} \bar{G}_0^{-1} \Phi -J^{\dagger}\Phi  +\frac{1}{2} \lambda  \left(\phi^* \phi \right)^2-\mu  \phi ^* \phi\right).
\end{equation}

The large-$\mathcal{N}$ expansion introduces $\mathcal{N}$ copies of the original systems with $\phi_n$, $n=1,2,\cdots,\mathcal{N}$, rescale the coupling constant as $\lambda/\mathcal{N}$, and sort the Feynman diagrams by powers of $1/\mathcal{N}$. For the single-species Bose gas studied here, we follow Ref.~\cite{ChienPRA12} and introduce 
an auxiliary field $\alpha$ representing $(\lambda/\mathcal{N})\sum_n\phi_n^*\phi_n$ via 
the following identity: 
\begin{eqnarray}
1&&=\int D
\alpha  \delta  \left(\alpha -(\lambda/\mathcal{N})  \sum_{n}\phi_n^* \phi_n \right)\nonumber\\
&&=\mathcal{C}\int D\alpha D \chi  e^{\frac{\chi  \left(\alpha -(\lambda/\mathcal{N})  \sum_{n}\phi_n^* \phi_n \right)}{\lambda }},
\label{eq:dirac_delta}
\end{eqnarray}
where $\mathcal{C}$ is a normalization factor and the $\chi$ integration runs parallel to the imaginary axis~\cite{Moshe200369}. 
After replacing $\sum_n\phi^*_n \phi_n $ by $\alpha\mathcal{N}/\lambda$, the Gaussian integration of $\phi_n$ and $\phi^*_n$ can be performed and one obtains
$Z[J,Y,K]=\int D\alpha  D \chi  e^{-S_{\text{eff}}}$,  
where we have introduced the sources $Y$ and $K$ for the auxiliary fields $\chi$ and $\alpha$, respectively.
To construct the leading-order theory, we only include up to the leading order of $1/\mathcal{N}$ in the effective action. Higher-order corrections of the $1/\mathcal{N}$ expansion can be constructed following Ref.~\cite{PhysRevA.83.053622}. After obtaining the leading-order $S_{\text{eff}}$, we set   $\mathcal{N}=1$ for single-component bosons. Then, 
\begin{eqnarray}
S_{\text{eff}}&=&
\int[dx]\left(-\frac{1}{2} J^{\dagger}  G_0 J-\frac{\alpha  \mu }{\lambda }-\frac{\alpha  \chi }{\lambda }+\frac{\alpha^2 }{2 \lambda }
-K\alpha - \right. \nonumber \\
& &\left. Y\chi  \right)+\frac{1}{2} \text{Tr} \ln G_0^{-1},
\end{eqnarray}
 $G_0^{-1}$ is defined as $G_0^{-1}\equiv \bar{G}_0^{-1}+diag(\chi,\chi)$.

The grand potential, which is also the generator of one-particle irreducible (1PI) graphs, can be obtained from a Legendre transform of the effective action \cite{nair2006quantum,zee2010quantum,PhysRevLett.105.240402} 
\begin{equation}
\Gamma[\phi_c,\phi_c^*,\chi_c,\alpha_c] =\int[dx]\left(J^{\dagger}\Phi _c +K \alpha _c+Y \chi _c\right)+S_{\text{eff}},
\end{equation}
where the subscript $c$ denotes the expectation values and will be dropped in the following. For static homogeneous fields, the effective potential is 
$V_{\text{eff}}=\Gamma /(\beta   \Omega)$.
Using the relation
$J=\int[dx]G_0^{-1}\Phi$  
we rewrite $V_{\text{eff}}$ in terms of the expectation value of $\Phi$ as
\begin{eqnarray}
V_{\text{eff}}&&=\frac{1}{2}\Phi ^{\dagger}G_0^{-1} \Phi-\frac{\alpha  \mu }{\lambda }-\frac{\alpha  \chi }{\lambda }+\frac{\alpha }{2 \lambda }+\nonumber\\
&&\sum _k \left(\frac{\omega _k}{2}+\frac{\ln \left(1-e^{-\beta  \omega _k}\right)}{\beta }\right),
\end{eqnarray}
where $ \omega _k=\epsilon _k+\chi$ and $\epsilon _k=\frac{k^2}{2 m}$. The last term in $V_{\text{eff}}$ comes from the trace log term, whose derivation is summarized in Appendix~\ref{app:TrLn}. For static homogeneous fields, only $\chi\phi^*\phi$ remains in the first term.

The expectation values of the fields can be found as the minimization conditions of the effective potential. From $\frac{\text{$\delta $V}_{\text{eff}}}{\delta \phi ^*}=0$, we get $\chi  \phi =0$,  which imposes the broken-symmetry (BEC) condition that when $\phi =0$ in the normal phase, $\chi$ is finite and $\chi=0$ in the broken symmetry phase when $\phi \neq 0$. The condition $\frac{\text{$\delta $V}_{\text{eff}}}{\delta \alpha }=0$ fixes the relation between $\chi$ and $\alpha$ by
\begin{equation}
\chi =\alpha -\mu.
\label{eq:1fchi}
\end{equation}
Since $V_{\text{eff}}$ is ultraviolet divergent, the theory needs to be 
renormalized. Ref.~\cite{ChienPRA12} detailed the renormalization of the leading-order large-$\mathcal{N}$ theory, and a brief summary is given in Appendix~\ref{app:Renormalization}. The renormalized effective potential is 
\begin{equation}
V_{\text{eff}}=-\frac{\alpha ^2}{2 \lambda }+\phi  \phi ^* (\alpha -\mu )+\sum _k \frac{\ln \left(1-e^{-\beta  \omega _k}\right)}{\beta }.
\label{eq:2fVReff}
\end{equation}
Here renormalized (physical) quantities are used. Importantly, $V_{\text{eff}}$ at the minimum is $-P$ from thermodynamics, where $P$ is the pressure of the system.

The equations of state can be derived from $\delta V_{\texttt{eff}}/\delta\phi^*=0$, $\frac{\delta  V_{\text{eff}}}{\delta \alpha}=0 $ and $-\frac{\delta  V_{\text{eff}}}{\delta  \mu }=\rho$. Explicitly, 
\begin{eqnarray}
(\alpha-\mu)\phi&=&0, \nonumber \\
\frac{\alpha }{\lambda }&=&\phi  \phi ^*+\sum _k n(\omega _k), \nonumber \\
\frac{\alpha }{\lambda }&=&\rho.
\label{eq:1frho}
\end{eqnarray}
Here $n(\omega_k)=[e^{\frac{\omega _k}{k_B T}}-1]^{-1}$ is the Bose distribution function. 
 In the normal phase, $\phi =0$ and $\omega _k=\epsilon _k+\alpha-\mu$. 
In the broken symmetry phase, $\phi$ is finite, so $\mu =\alpha $ and $\omega _k=\epsilon _k$. By the $U(1)$ symmetry of $\phi$, we can choose $\phi$ to be real in the broken symmetry phase and associate $\phi=\sqrt{\rho_c}$ with the condensate density $\rho_c$.

\subsection{LOAF theory}
The leading-order large-$\mathcal{N}$ theory is a conserving theory with a consistent thermodynamic free energy. It also shows a second-order BEC transition. However, one major issue with the leading-order large-$\mathcal{N}$ theory is its inconsistency with the Bogoliubov theory of weakly interacting bosons at zero temperature \cite{ChienPRA12}. Given theoretical and experimental support of the Bogoliubov dispersion of weakly interacting bosons near zero temperature \cite{RevModPhys.76.599,PhysRevLett.88.120407}, the leading-order large-$\mathcal{N}$ theory needs further improvement. 
The main reason of the inconsistency is because the anomalous density representing pairing correlations, $A=\lambda\langle\phi\phi\rangle$, is included in the Bogoliubov theory but not in the leading-order large-$\mathcal{N}$ theory. By including the normal density composite field $\chi=\sqrt{2}\lambda\langle\phi^*\phi\rangle$ and $A$,  a similar large-$\mathcal{N}$ expansion leads to the LOAF theory \cite{PhysRevLett.105.240402,PhysRevA.83.053622,PhysRevA.84.023603,PhysRevA.85.023631,PhysRevLett.105.240402}, which is fully consistent with the Bogoliubov theory in the weakly interacting regime. A brief summary of the derivation of the LOAF theory is in Appendix~\ref{app:LOAF}.

The regularized LOAF effective potential is 
\begin{eqnarray}
V_{\text{eff}}
&&=\chi ^{\prime} \phi ^* \phi -\frac{1}{2} A^* \left(\phi ^*\right)^2-\frac{A \phi ^2}{2}-\frac{\left(\chi ^{\prime}+\mu \right)^2}{4 \lambda }+\frac{A A^*}{2 \lambda }+\nonumber\\
&&\sum _k \left(\frac{1}{2} \left(\omega _k-\epsilon _k-\chi ^{\prime}+\frac{A A^*}{2 \epsilon _k}\right)+\right.\nonumber\\
&&\left. T \ln \left(1-e^{-\frac{\omega _k}{T}}\right)\right),
\end{eqnarray}
where $\chi^{\prime} \equiv \sqrt{2}\chi-\mu$.
From the minimization conditions,$-\frac{\text{$\delta $V}_{\text{eff}}}{\delta \mu }
=\rho$ and $\frac{\text{$\delta $V}_{\text{eff}}}{\delta \phi }=\frac{\text{$\delta $V}_{\text{eff}}}{\delta \chi ^{\prime}}=\frac{\text{$\delta $V}_{\text{eff}}}{\text{$\delta $A}^*}=0$, we arrive at the equations of state 
\begin{eqnarray}
\rho
=\frac{\mu +\chi ^{\prime}}{2 \lambda },
\label{eq:2fmu}
\end{eqnarray}
\begin{eqnarray}
0
=\phi ^* \chi ^{\prime}-A \phi,
\label{eq:2fchi}
\end{eqnarray}
\begin{eqnarray}
0
&&=- \frac{\mu +\chi ^{\prime}}{2 \lambda }+\rho_c +\nonumber\\
&&\sum _k \left(\frac{\left(\epsilon _k+\chi ^{\prime}\right) \left(1+2n(\omega_k)\right)}{2 \omega _k}-\frac{1}{2}\right).
\label{eq:2frho}
\end{eqnarray}
\begin{eqnarray}
0
&&=
\frac{A}{\lambda}- \rho_c- A \sum _k \left(\frac{1+2 n(\omega_k)}{2 \omega _k}-\frac{1}{2 \epsilon _k}\right).
\label{eq:2fA}
\end{eqnarray}
In the BEC phase, we use the $U(1)$ symmetry to choose the expectation value of $\phi$ to be real and equal to $\sqrt{\rho_c}$ with $\rho_c$ being the condensate density.

The leading-order large-$\mathcal{N}$ theory has two phases: Above $T>T_c$ it gives a normal phase, where the condensate $\phi=0$ but the composite field $\chi>0$ playing the role of the chemical potential. Below $T_c$ it is a broken symmetry phase corresponding to BEC, where the condensate $\phi>0$ and $\chi=0$~\cite{ChienPRA12}. The LOAF theory, on the other hand, predicts three possible phases: At high $T$ it is a normal phase, where both the condensate $\phi$ and the anomalous density $A$ vanish. The composite field $\chi^{\prime}>0$ is related to the chemical potential. Below $T_c$ it is a broken symmetry (BEC) phase with $\phi>0$ and $A>0$. The composite field $\chi$ is related to $A$ according to Eq.~\eqref{eq:2fchi}. Interestingly, there is an intermediate-temperature superfluid phase in the regime $T_c<T<T^*$, where the condensate vanishes $\phi=0$ but the anomalous density remains finite, $A>0$. The finite $A$ gives rise to a finite superfluid density as derived in Ref.~\cite{PhysRevA.85.023631}. In the intermediate-temperature superfluid regime, the two composite fields $\chi$ and $A$ are different and need to be determined from a set of coupled equations.

\subsection{Leading-order $T_c$ shift}
As mentioned before, the leading-order $T_c$ shift of a trapped Bose gas has been evaluated in Refs~\cite{PhysRevA.54.R4633} using the Popov approximation with the LDA. To compare with the experimental results in Ref.~\cite{PhysRevLett.106.250403}, we introduce the thermal de Broglie wavelength of a trapped noninteracting Bose gas with the same total particle number $N$ at its critical temperature, $\lambda_0=\sqrt{\frac{2 \pi \hbar ^2}{m k_B T_c^0}}$. The leading order in the Popov approximation is then 
\begin{equation}
\frac{\delta T_c}{T_c^0}=-3.4260\frac{a}{\lambda_0}.
\end{equation}

For the leading-order large-$\mathcal{N}$ theory, the dispersion above $T_c$ is $\omega_k(r)=k^2+\chi(r)$, where $\chi(r)=\lambda\rho(r)-\mu(r)$. When compared to the dispersion of the Popov approximation, the dispersion is almost identical if $\lambda$ in the Popov approximation is replaced by $\lambda/2$. This substitution leads to the $T_c$ shift in the leading-order large-$\mathcal{N}$ theory as
\begin{equation}
\frac{\delta T_c}{T_c^0}=-1.7130\frac{a}{\lambda_0}.
\end{equation}
In the LOAF theory, the two critical temperatures $T^*$ and $T_c$ merge in the weakly interacting regime \cite{PhysRevA.85.023631}. Therefore, for analytic calculations we use the shift of $T^*$ as a proxy to estimate the $T_c$ shift of a trapped gas in the weak interaction regime. From the discussion of the LOAF theory in Appendix~\ref{app:LOAF}, the local composite fields are $\chi^{\prime}(r)=-\mu(r)+2\lambda\rho(r)$ by Eq.~\eqref{eq:1fchi} and $A(r)=0$ at $T^*$, so Eq.~\eqref{eq:2frho} reduces to $\rho(r)=\sum _k n(\omega_k)$, where $\omega_k(r)=k^2+V(r)+2\lambda\rho(r)-\mu_0$. This dispersion is identical to the Popov approximation to the lowest order in the coupling constant \cite{PhysRevA.54.R4633}. Following a similar calculation, 
\begin{equation}
\frac{\delta T_c}{T_c^0}=-3.4260\frac{a}{\lambda_0}.
\end{equation}
Thus the leading order result of the LOAF theory agrees with the Popov theory and experimental data \cite{PhysRevA.54.R4633,PhysRevLett.106.250403}. Our numerical results using the LDA agree well with the leading-order estimations presented here.

\section{large-$\mathcal{N}$ based theories for trapped Bose gases}\label{sec:LargeNLDA}
The large-$\mathcal{N}$ based theories can be formulated with the LDA, where the trap potential $V(r)$ is grouped with the chemical potential $\mu_0$ and the local chemical potential $\mu(r)=\mu_0-V(r)$ is introduced. Then we search for a solution consistent with the profile of $\mu(r)$ with a given $N$.
 To find the density profile $\rho(r)$ from large-$\mathcal{N}$ based theories numerically, the following procedures have been implemented. Since the total particle number $N$ and temperature $T$ are given, one needs to find the chemical potential $\mu_0$ satisfying Eq.~\eqref{eq:number}. 
This also implies that $\mu_0$ is a function of $T$ and the coupling constant.
For harmonically trapped systems, the following units are introduced.
$a_H\equiv\sqrt{\frac{\hbar }{m \omega }}$ and 
$E_0\equiv \frac{\hbar ^2}{2 m a_H^2}$. 
 Then 
$\frac{\lambda_0 }{a_H}=\sqrt{2 \pi } \left(\frac{\zeta (3)}{N}\right)^{1/6}$. 
This allows us to use the following dimensionless quantities. 
$\frac{\epsilon _k}{E_0}=\left(k a_H\right)^2$, 
$\frac{V(r)}{E_0}=\left(\frac{r}{a_H}\right)^2$, and
$\frac{k_B T}{E_0}$. 
Moreover, the (renormalized) coupling constant is related to the two-body $s$-wave scattering length by
\begin{equation}
\frac{\lambda }{E_0 a_H^3}=8\pi\frac{a}{a_H}.
\end{equation}

\subsection{Leading-order large-$\mathcal{N}$ theory with LDA}
We begin with the leading-order large-$\mathcal{N}$ theory with the LDA. At given $T$ and $a$, $\mu_0$ should satisfy Eq.~\eqref{eq:number} with a density profile $\rho(r)$ determined from the equations of state on a grid discretizing the geometry. We have chosen the grid size small enough that further reductions of the size do not change our results. Initially, a trial value of $\mu_0$ is guessed and we find the corresponding $\rho(r)$. If BEC is present, we need to locate the size of the condensate. This is equivalent to finding a critical radius $r_c$ where $\rho_c(r_c)=0$. At $r=r_c$ the condition $\chi(r_c)=\rho_c(r_c)=0$ can be used in Eq.~\eqref{eq:1fchi} and Eq.~\eqref{eq:1frho} to obtain $r_c
=\sqrt{(\mu _0-\left(\frac{ m k_B T}{2 \pi  \hbar ^2}\right)^{3/2} \zeta \left(\frac{3}{2}\right) \lambda)(\frac{2}{m \omega ^2})}$.

Once $r_c$ is located, the density profile $\rho(r)$ can be constructed with the information of $T$ and $\mu(r)$. If $T<T_c$, there is a condensate within $r<r_c$, whose condensate density can be found from $\rho_c(r)=\rho(r)-\rho_T(r)$. In this region $\chi(r)=0$ due to the BEC condition, so $\rho(r)=\mu(r)/\lambda$ by Eq.~\eqref{eq:1fchi} and Eq.~\eqref{eq:1frho} gives us $\rho_T(r)=\left(\frac{k_B T}{4 \pi  E_0}\right){}^{3/2} \zeta \left(\frac{3}{2}\right)a_H^3$.  
Outside the condensate region ($r>r_c$), $\rho_c(r)=0$ and one can solve Eq.~\eqref{eq:1frho} to obtain $\rho(r)$. After $\rho(r)$ in the whole trap is found, $\mu_0$ can be evaluated by iteratively solving Eq.~\eqref{eq:number} by treating $\mu_0$ as a function of $N$, $T$, and $a$.

Above $T_c$ there is no condensate ($\rho_c(r)=0$), and a similar procedure leads to $\mu_0$ and $\rho(r)$ as well. To find $T_c$, we tune the temperature so that $r_c=0$. This is the temperature when the condensate is about to emerge. The relation between $T_c$ and $\mu_0(T_c)$ are fixed by the expression of $r_c=0$, and $T_c$ can be found by iteratively search for the solution that satisfies Eq.~\eqref{eq:number} with given $N$ and $a$.

\subsection{LOAF theory with LDA}
As mentioned before, the LOAF theory for a uniform interacting Bose gas exhibits a richer phase diagram with three distinct phases. 
For a trapped Bose gas below $T_c$, there is a condensate at the center with $\rho_c(r)>0$ and $A(r)=\chi(r)>0$. The condensate vanishes at $r_c$, where $\rho_c(r\ge r_c)=0$ but $A(r_c)$ can still be finite. The anomalous density vanishes at $r=r^*$, and outside $r^*$ the system is normal with $\rho_c(r)=0$ and $A(r)=0$. 

To find the density profile and $\mu_0$ with given $N$, $T$, and $a$, we solve Eq.~\eqref{eq:number} iteratively with the following procedures. The initial value of $\mu_0$ is guessed and we map out the corresponding $\rho(r)$. Next we need to locate $r_c$ where $\rho_c(r_c)=0$. From $A(r_c)=\chi(r_c)$ and $\rho(r_c)=0$, Eq.~\eqref{eq:2fA} and Eq.~\eqref{eq:2frho} allow us to determine $\chi(r_c)$ and $\mu(r_c)$. Then $r_c$ can be inferred from $\mu_0-\mu(r_c)$. When $r>r_c$, the anomalous density $A(r)$ should decay to zero at $r=r^*$. Using $A(r\rightarrow r^*)\rightarrow 0$ in Eq.~\eqref{eq:2fA}, one can find $\chi(r^*)$, which can be used in Eq.~\eqref{eq:2frho} to find $\mu(r^*)$. Then $r^*$ is inferred from $\mu_0-\mu(r^*)$. 

After determining $r_c$ and $r^*$, we can map out the whole density profile. To find $\rho(r)$ in the condensate region, we set $A(r)=\chi(r)$ with a finite $\rho_c(r)$. Multiplying Eq.~\eqref{eq:2frho} by $\lambda$ and subtracting Eq.~\eqref{eq:2fA} lead to an equation for $\chi(r)$. After solving for $\chi(r)$, we can get $\rho(r)$ from Eq~\eqref{eq:2fmu} with $\mu(r)$ and $\chi(r)$. In the region between $r_c$ and $r^*$, $A(r)$ and $\chi(r)$ are found by solving Eq.~\eqref{eq:2frho} and Eq.~\eqref{eq:2fA} simultaneously. Once $\chi(r)$ is found, $\rho(r)$ is again obtained by Eq.~\eqref{eq:2fmu}. Outside $r^*$, $A(r)=0$ and only Eq.~\eqref{eq:2frho} needs to be solved, which will give us $\chi(r)$ and thus $\rho(r)$. After $\rho(r)$ is obtained in all regions, $\mu_0$ is solved iteratively by Eq.~\eqref{eq:number}, where the integral is split over different regions determined by $r_c$ and $r^*$.

The critical temperature corresponds to a density profile with $r_c=0$ when the condensate at the center is about to emerge. The condition $r_c=0$ fixes the relation between $T_c$ and $\mu_0$ at the center by Eq.~\eqref{eq:2fA} with $\rho_c(r_c)=0$ and $\chi(r_c)=A(r_c)$, and we only need to find $r^*$ and the whole density profile. Then by solving Eq.~\eqref{eq:number} iteratively we obtain $T_c$ from the LOAF theory with the LDA.

\section{Results and Discussions}\label{sec:Results}
Since the thermodynamic limit has been taken in each slice of the LDA, the number-fixing procedure, Eq.~\eqref{eq:number}, serves to fix the units. In our calculations we set $N=1000$, which corresponds to $\frac{\lambda_0 }{a_H}=0.8173$. By choosing a different value of $N$ and scaling the units accordingly, the coefficients in the expression of $T_c$ shift remain the same.

Using the experimental data from Ref.~\cite{PhysRevLett.92.030405} and subtracting finite-size effects, Ref.~\cite{PhysRevLett.96.060404} showed a $T_c$-shift curve as a function of $a/\lambda_0$. Only linear and quadratic terms are used over the range of experimental data and corrections from the logarithmic term suggested in Ref.~\cite{PhysRevA.64.053609} was not found. We follow the clue and did not include the logarithmic term when extracting the functional form of our results.

\begin{figure}
\includegraphics[width =3.4in]{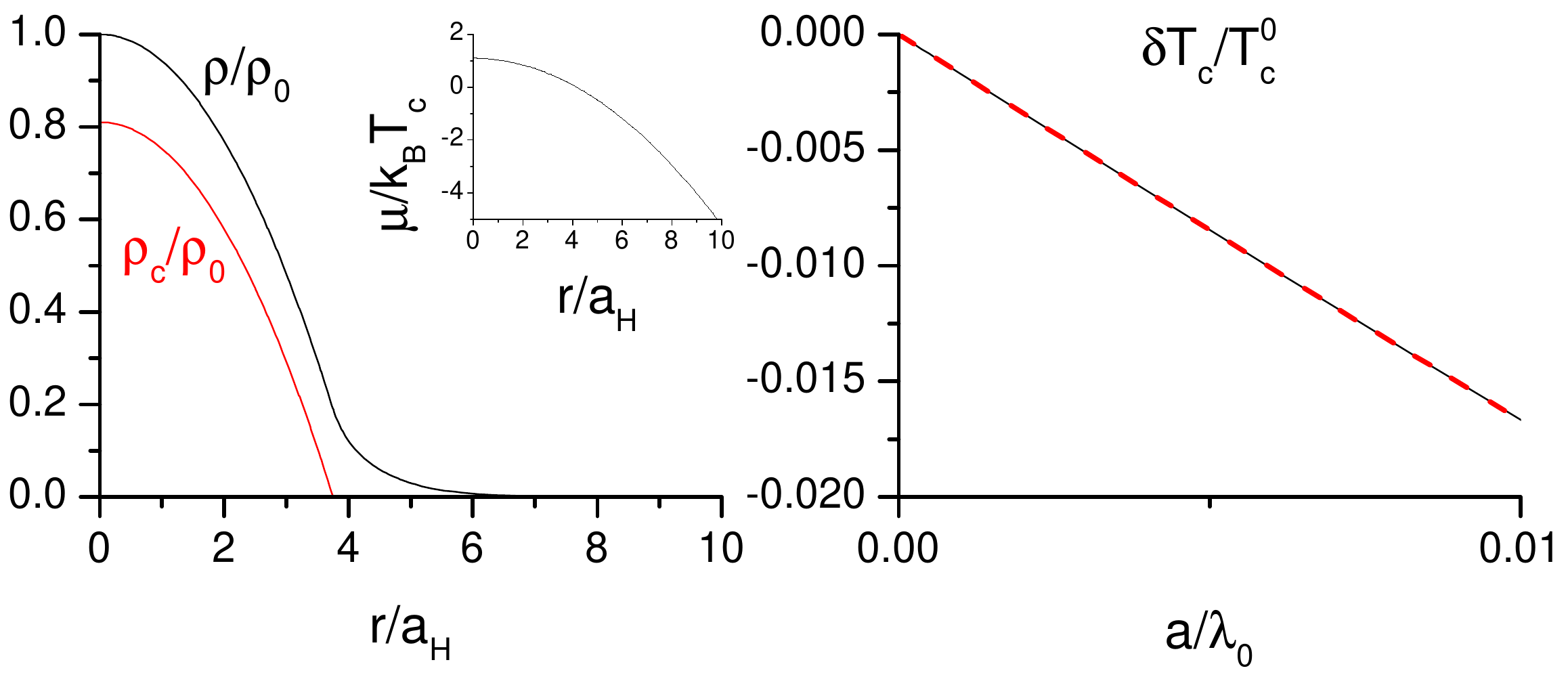}
\caption{(Color online) Leading-order large-$\mathcal{N}$ theory with the LDA. (Left) Density profile with  $a/a_H=0.1$ and $T/T_c=0.5$. Here $\rho_0$ is the density at the trap center. The inset shows the local chemical potential. (Right) $T_c$ shift (black solid line) with a fit (red dashed line). The functional form is shown in Eq.~\eqref{eq:LNTcFit}.  Here $\frac{\lambda_0 }{a_H}=\sqrt{2 \pi } \left(\frac{\zeta (3)}{N}\right)^{1/6}$ with $N=1000$.} 
\label{fig:1field}
\end{figure}

The $T_c$ shift from the leading-order large-$\mathcal{N}$ theory with the LDA is presented in Figure~\ref{fig:1field}. The curve is almost linear with $a/\lambda_0$ and has a very small curvature. A fitting of the curve gives 
\begin{equation}\label{eq:LNTcFit}
\frac{\delta T_c}{T_c^0}=-1.71\frac{a}{\lambda_0}+4.55\left(\frac{a}{\lambda_0}\right)^2. 
\end{equation}
When compared to the experimental data from Ref.~\cite{PhysRevLett.106.250403}, the coefficient of the leading-order term is only half of the experimental value and the coefficient of the next-order term is even farther away.

\begin{figure}
\includegraphics[width = 3.4in]{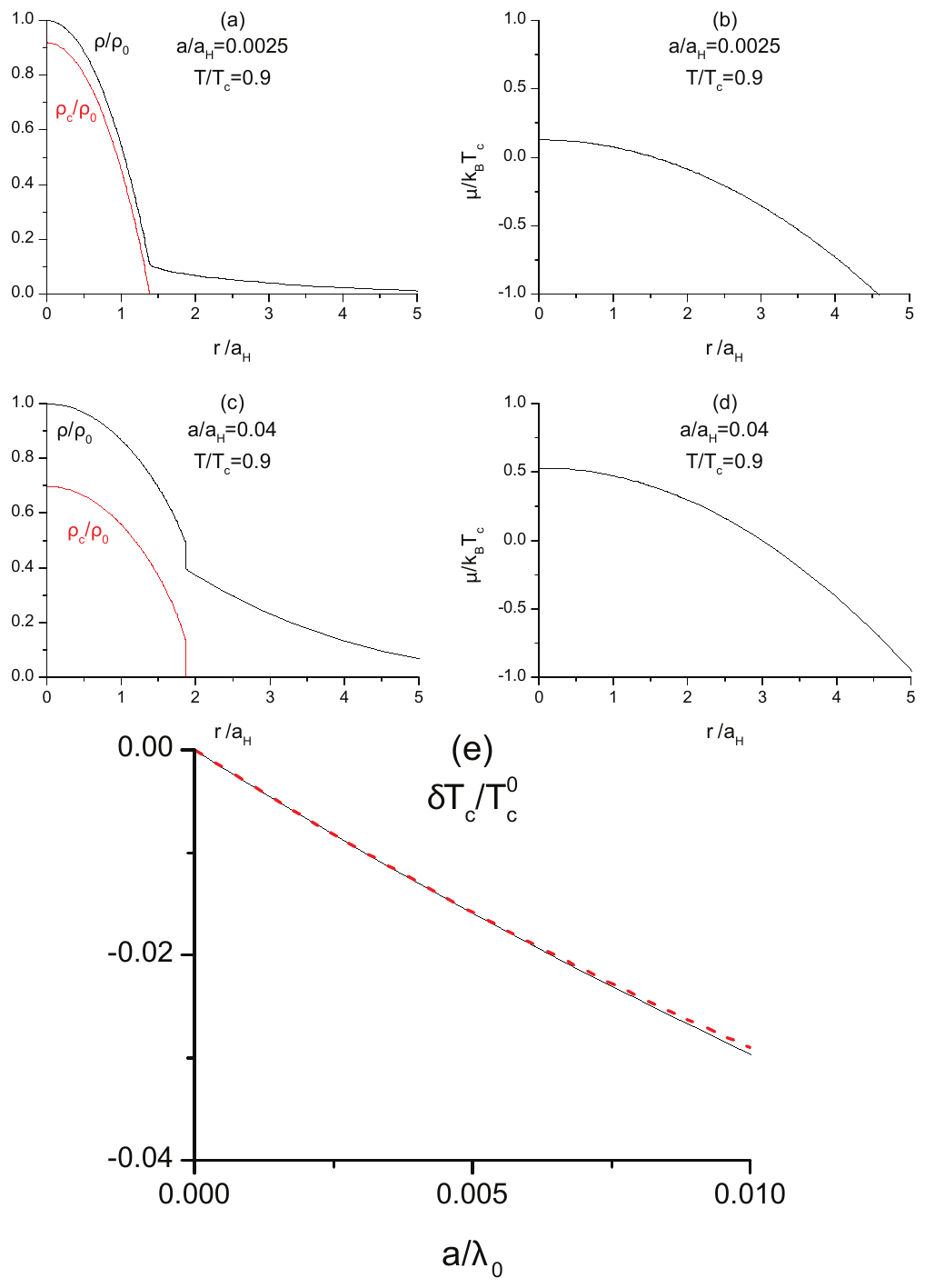}
\caption{(Color online) LOAF theory with LDA. (a) and (c) show the trap profiles of the density and condensate at $T/T_c=0.9$ for $a/a_H=0.0025$ and $a/a_H=0.04$. The corresponding local chemical potentials are shown in (b) and (d), respectively. Discontinuities are observable at the boundary separating superfluid and normal phases. (e) $T_c$ shift (black solid line) with a fit (red dashed line). The functional form is shown in Eq.~\eqref{eq:LOAFTcFit}. Here $N=1000$. 
\label{fig:2field}}
\end{figure}

The $T_c$ shift from the LOAF theory with the LDA is shown in Figure~\ref{fig:2field}. The density profile exhibit a density discontinuity at the boundary of superfluid and normal phases, and we will comment on this behavior later on. For the weakly interacting regime, a fitting of the $T_c$ shift gives 
\begin{equation}\label{eq:LOAFTcFit}
 \frac{\delta T_c}{T_c^0}=-3.42\frac{a}{\lambda_0}+52.00\left(\frac{a}{\lambda_0}\right)^2. 
\end{equation}
When compared to the expression extracted from the experimental data of Ref.~\cite{PhysRevLett.106.250403}, we found excellent agreements for the leading-order as well as the next-order terms. Due to difficulties of formulating mean-field theories of trapped interacting Bose gases, to our knowledge a theoretical evaluation of the quadratic term in the $T_c$ shift has not been available. The LOAF theory with the LDA thus may serve as a manageable mean-field theory for describing the $T_c$ shift of trapped interacting Bose gases.

A summary of the coefficients of $T_c$ shift from large-$\mathcal{N}$ based theories and the experimental data of Ref.~\cite{PhysRevLett.106.250403} is given in Table~\ref{tab:comparison}. One can see that by introducing the anomalous density $A=\lambda\langle\phi\phi\rangle$ originally incorporated in the Bogoliubov theory, the predictions of the $T_c$ shift improve substantially from the leading-order large-$\mathcal{N}$ theory to the LOAF theory. When compared to previous theoretical studies using the Popov theory \cite{PhysRevA.54.R4633} allowing for an extraction of the leading-term coefficient, the large-$\mathcal{N}$ base theories allow us to fit the functional form of the $T_c$ shift with higher orders and analyze the trap density profiles. Moreover, the agreement between the LOAF theory and experimental data from Ref.~\cite{PhysRevLett.106.250403} does not require the logarithmic term from Ref.~\cite{PhysRevA.64.053609}.
\begin{table}
    \begin{tabular}{| l | l | l |}
    \hline
     & $b_1$ & $b_2$ \\ \hline
    Leading-order large-$\mathcal{N}$ & -1.71 & 4.55 \\ \hline 
    LOAF & - 3.42 & 52.00  \\ \hline
    Phenomenological model & - 3.38 & 30.35  \\ \hline
    Experiment~\cite{PhysRevLett.106.250403} & $-3.5\pm0.3$ & $46\pm5$  \\
    \hline
           \end{tabular}
    \caption{Comparsion of theoretical and experimental results. The $T_c$ shift has the form $\frac{\text{$\delta $T}_c}{T_c^0}=b_1\frac{a}{\lambda_0 }+b_2 \left(\frac{a}{\lambda_0 }\right)^2$. The phenomenological model predicts a cubic term with a coefficient $-152.5$. Choosing different values of $N$ only scales the units, and the coefficients shown here do not change.}
            \label{tab:comparison}
\end{table}

\subsection{Incompatible behavior of LOAF theory with LDA}
As shown in Fig.~\ref{fig:2field}, the LOAF theory with the LDA exhibits an observable density discontinuity between the superfluid and normal phases. The discontinuity increases as $a$ increases. We caution that, although density discontinuities were also found in the Popov theory with the LDA \cite{BF02396737}, the origins of the discontinuities are very different. For the Popov theory, a discontinuous first-order transition already emerges in a homogeneous interacting Bose gas \cite{RevModPhys.76.599,BF02396737,PhysRevLett.105.240402}. In contrast, the LOAF theory predicts a smooth second-order transition for homogeneous interacting Bose gases \cite{PhysRevLett.106.250403}. The discontinuity in the trap profile of the LOAF theory comes from incompatibility of the theory with the LDA requiring $\mu(r)$ to decrease quadratically in a harmonic trap. The incompatibility will be elaborated, and here we emphasize that the leading-order large-$\mathcal{N}$ theory does not suffer from any density discontinuity even in the stronger interaction regime (illustrated in Figure~\ref{fig:1field}), but its predictions of the $T_c$ shift do not agree quantitatively with the experimental data of Ref.~\cite{PhysRevLett.106.250403} as summarized in Table~\ref{tab:comparison}.

The density discontinuity of the LOAF theory can be analyzed as follows. For $r\ge r_c$, $\rho_c=0$ in Eq.~\eqref{eq:2fA}. Thus, at $r_c$ and $r^*$ the following equation has to be satisfied with different energy dispersions.
\begin{equation}
1
=
\lambda   \sum _k \left(\frac{1+2 n(\omega_k)}{2 \omega _k}-\frac{1}{2 \epsilon _k}\right).
\label{eq:A_c}
\end{equation}
The dispersion at $r=r_c$ is $\omega_{kc}=\sqrt{\epsilon_k (\epsilon_k+2\chi_c^{\prime})}$ with $\chi_c^{\prime}$ denoting the value of $\chi^{\prime}$ at $r_c$. At $r=r^*$,  $\omega_{k}^*=\epsilon_k+\chi^{\prime *}$ with $\chi^{\prime *}$ denoting the value of $\chi^{\prime}$ at $r^*$. 
If $\chi_c^{\prime}$ is less than $\chi^{\prime *}$, $\omega_{k}^*$ will be greater than $\omega_{kc}$ for any $k$, so Eq.~\eqref{eq:A_c} cannot be satisfied by both dispersions. This problem can be circumvented by making $\omega_{k}^*$ less than $\omega_{kc}$ at small $k$ so that Eq.~\eqref{eq:A_c} can be satisfied by both dispersions. This requires $\chi_c^{\prime}$ to be greater than $\chi_{k}^*$. However, if $\lambda\rho(r)$ does not decrease fast enough as $r$ increases, $\chi^{\prime}(r)=-(\mu_0-\frac{1}{2}m\omega^2r^2)+2\lambda\rho(r)$ may be an increasing function. When it happens, $r^*>r_c$ does not exist.

One solution is to force the LDA form of $\mu(r)$ and connect the superfluid and normal phases similar to the Maxwell construction. Then the solution exhibits a jump of $A(r)$ to zero at $r_c$. Such a discontinuity in $A(r)$ then cause a discontinuity in the density profile according to Eq.~\eqref{eq:2fmu}, which is observable in Fig.~\ref{fig:2field} (c).  The incompatibility with the LDA is also hinted by the behavior of $\mu$ as a function of $T$. In the LOAF theory of homogeneous Bose gases, $\mu(T)$ can be non-monotonic as $T$  increases~\cite{PhysRevLett.105.240402}. However, for a trapped Bose gas in the LDA, the local chemical potential $\mu(r)=\mu_0-\frac{1}{2}m\omega^2r^2$ should decrease quadratically as $r$ increases. Since the particle density decreases with $r$ and the local temperature scale $T(r)$ is determined by the local density, the temperature ratio $T/T(r)$ increases with $r$. Therefore, when the interaction is too strong and $\mu(T)$ exhibits prominent non-monotonicity, the LOAF theory cannot be pieced together in the LDA. We remark that the leading-order large-$\mathcal{N}$ theory does not have such incompatibility with the LDA.

\subsection{Phenomenological model}\label{sec:phe}
The incompatibility of the LOAF theory with the LDA leads us to contemplate possible alternatives. One may re-derive the whole theory in real space with inhomogeneity. The fields are no longer uniform and the equations of state will be coupled differential equations. Solving the equations is not only numerically demanding, but also loses transparency in explaining the underlying physics. Here we explore a phenomenological alternative by requiring that the local anomalous density $A(r)$, instead of the local chemical potential $\mu(r)$, decays with a quadratic form. Explicitly, the condition $A(r)=A_0-E_0(r/a_H)^2$ is imposed and the anomalous density at the trap center, $A_0$, needs to be solved iteratively.

To obtain the density profile and corresponding chemical potential, $A_0$ is guessed initially and we use $A(r)$ to map out $\rho(r)$. Since $A(r)$ has to vanish at $r=r^*$, we have $r^*/a_H=\sqrt{A_0/E_0}$. To find $r_c$, we first find $A(r_c)$ by solving Eq.~\eqref{eq:2fA} at $r_c$ with $\chi(r_c)=A(r_c)$. Then $r_c$ is found by $r_c/a_H=\sqrt{A_0-A(r_c)}$. Moreover, at $r_c$ one has $\chi(r_c)=A(r_c) >0$, so $r_c$ cannot be greater than $r^*$. In the region $r<r_c$, $\chi(r)=A(r)$ and $\mu(r)$ is obtained by multiplying Eq.~\eqref{eq:2frho} by $\lambda$ and subtracting Eq.~\eqref{eq:2fA}. Then the resulting equation can be solved to give $\chi(r)$. In the region $r_c<r<r^*$, $\chi(r)$ is obtained by solving Eq.~\eqref{eq:2fA} and then Eq.~\eqref{eq:2frho} is used to obtain $\mu(r)$. For $r^*<r$, $A(r)=0$ and we use the LDA for local chemical potential $\mu(r)=\mu(r^*)-E_0\left(\left(r-r^*\right)/a_H\right)^2$ to finish the computation. Outside $r^*$, $\chi(r)$ is obtained by solving Eq.~\eqref{eq:2frho}. After $\chi(r)$ and $\mu(r)$ are found in all regions, $\rho(r)$ can be inferred by Eq.~\eqref{eq:2fmu}. Then $A_0$ is found by iteratively solving Eq.~\eqref{eq:number} with given $T$ and $a$.

\begin{figure}
	\includegraphics[width = 3.4in]{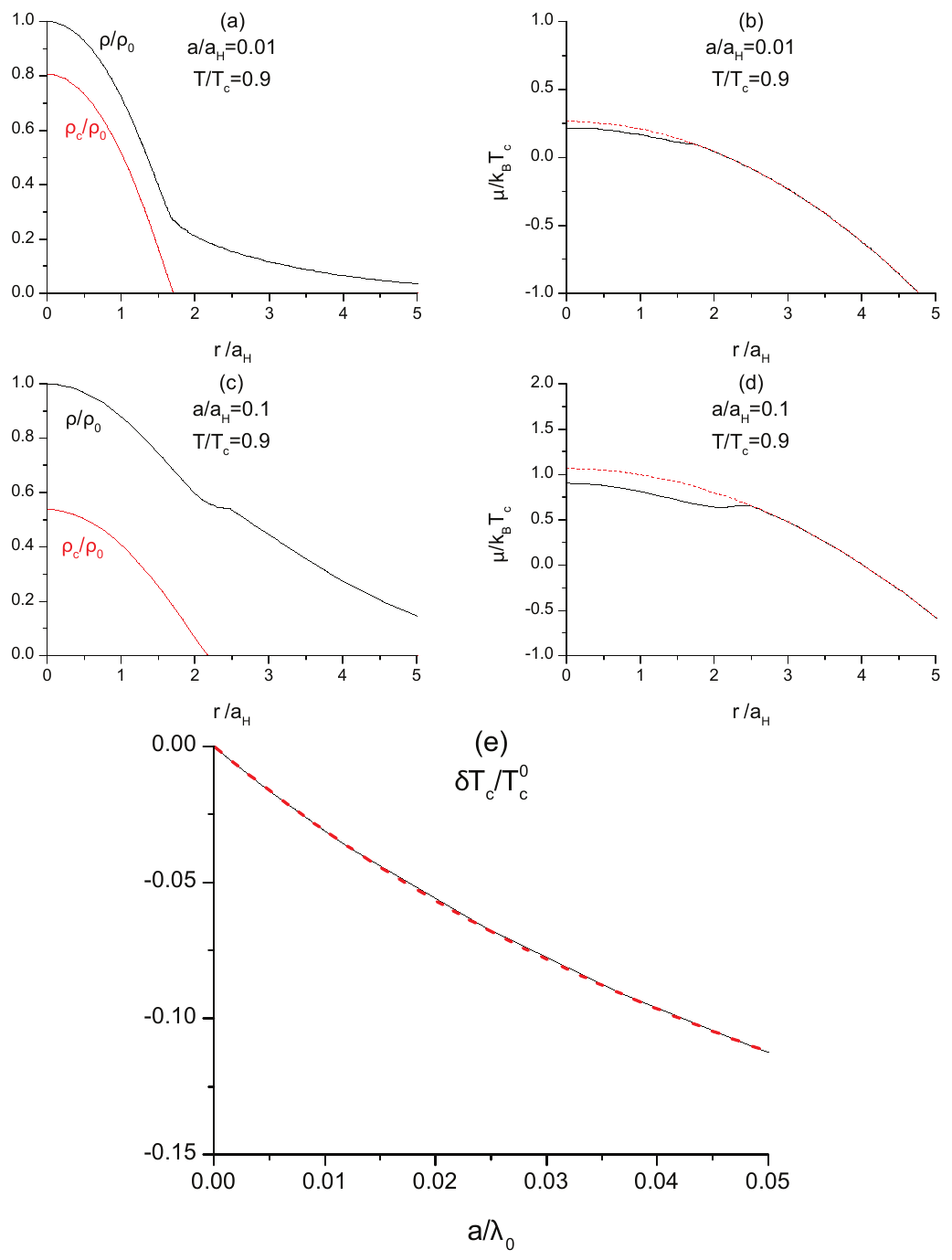}
	\caption{(Color online) The phenomenological model discussed in Sec.~\ref{sec:phe}. (a) and (c): Density profiles for  $a/a_H=0.01$ with $T=0.9T_c$ and for $a/a_H=0.1$ with $T=0.9T_c$, respectively. Note that $T_c$ shifts with interaction. The condensate density $\rho_c$ and total density $\rho$ are normalized by the density at the trap center, $\rho_0$. The corresponding local chemical potentials are shown in (b) and (d), respectively. The black solid lines are $\mu(r)$ obtained from the phenomenological model and the red dotted lines are extrapolations using the LDA. Here $N=1000$. (e) $T_c$ shift (black solid line) and a fit (red dashed line). The functional form is shown in Eq.~\eqref{eq:funnyTcFit}.}
	\label{fig:funny_model}
\end{figure}

As shown in Figure~\ref{fig:funny_model} (a)-(d), both the density profile and local chemical potential of the phenomenological model are continuous in the whole trap. However, the local chemical potential clearly exhibits a deviation from the conventional LDA. To contrast the difference, the red dotted line shows an extrapolation according to the LDA with $\mu(r)=\mu_0^2-E_0(r/a_H)^2$, where $\mu_0$ is calculated to match the chemical potential of this phenomenological model outside $r^*$. The non-monotonic local chemical potential as a function of $r$ confirms the incompatibility of the LOAF theory with the LDA, and with a simple reformulation of the local anomalous density $A(r)$ we restore continuity to the trapped system.

Figure~\ref{fig:funny_model} further illustrates the phenomenological model for different interaction strength close to $T_c$. Apparently, the deviation from the conventional LDA becomes more prominent as the interaction increases. The deviation is understandable because interactions should lead to corrections of the chemical potential in a many-body system, and the simple assumption of grouping $\mu$ and the trapping potential $V(r)$ at the bare level may no longer hold. We caution that the model with an LDA form of $A(r)$ is purely phenomenological, and a full treatment of the LOAF theory with inhomogeneous fields using numerical methods will eventually replace the phenomenological model and the LDA.

The $T_c$ shift predicted by the phenomenological model is shown in Fig.~\ref{fig:funny_model} (e). In contrast to the LOAF theory with the LDA, we did not find any density jump in the trap profile of the phenomenological model. By fitting the curve in a broader range of interaction strength, we found the functional form of the $T_c$ shift as
\begin{equation}\label{eq:funnyTcFit}
\frac{\delta T_c}{T_c^0}=-3.382\frac{a}{\lambda_0}+30.35\left(\frac{a}{\lambda_0}\right)^2-152.5 \left(\frac{a}{\lambda_0}\right)^3. 
\end{equation}
The coefficient of the cubic term serves as a prediction for future experimental measurements in stronger interaction regime.

\section{Conclusion}\label{sec:conclusion}
Two large-$\mathcal{N}$ based theories, the leading-order large-$\mathcal{N}$ theory and the LOAF theory, implemented with the LDA capture interesting physics of the $T_c$ shift of harmonically trapped interacting Bose gases. The leading-order large-$\mathcal{N}$ theory produces continuous trap profiles in weak and intermediate interaction regimes, but quantitative comparisons with available experimental data do not agree quantitatively. On the other hand, the LOAF theory with the LDA predicts a functional form with  higher-order corrections agreeing well with experimental data. The LOAF theory also signals  beyond-LDA behavior in the trap profile. The proposed phenomenological model based on the LOAF theory may serve as a useful patch for studying trapped Bose gases when a full analysis of inhomogeneous fields remains a great challenge.

We thank Fred Cooper and Kevin Mitchell for stimulating discussions.

\appendix
\section{Details of large-$\mathcal{N}$ based theories}
\subsection{Calculation of $Tr \ln \bar{G}_0^{-1}$}\label{app:TrLn}
To evaluate the $Tr \ln \bar{G}_0^{-1}$ term in the effective action, we define $I(s)$ as 
\begin{eqnarray}
I (s)
&&\equiv\frac{1}{2}\int d^4x \int \frac{d^3k}{\left(2\pi\right)^3}\nonumber\\
&&T \sum _n \log \left(\omega _n^2+\omega_k^2+s\right),
\end{eqnarray}
where $\omega_n=2n\pi/\beta$ is the bosonic Matsubara frequency and $\omega_k$ is the energy dispersion. It can be shown that $I(0)=\frac{1}{2} \text{Tr}\left[\ln \left(\bar{G}_0^{-1}\right)\right]$ \cite{nair2006quantum,zee2010quantum}.
Then, 
\begin{eqnarray}
\frac{\partial I(s)}{\partial s}&&=\frac{1}{2}\int d^4x \int \frac{d^3k}{\left(2\pi\right)^3}T \sum _n \frac{1}{\omega _n^2+\omega_k^2+s}\nonumber\\
&&=\frac{1}{2}\int d^4x \int \frac{d^3k}{\left(2\pi\right)^3}T\nonumber\\
&&\oint\frac{1}{\left((2 \pi  T z)^2+\omega_k^2+s\right) \left(e^{2 \pi  i z}-1\right)}
\end{eqnarray}
where the contour encircles the real axis counterclockwise. We deform the contour on the upper complex plane to encircle the upper half complex plane and the contour on the lower complex plane to encircle the lower half complex plane, both clockwise. 
Taking the residues on the imaginary axis, $2\pi T z=\pm i \sqrt{\omega_k^2+s}$, one obtains  
\begin{eqnarray}
\frac{\partial I(s)}{\partial s}&&=\frac{1}{2}\int d^4x \int \frac{d^3k}{\left(2\pi\right)^3}\nonumber\\
&&\left(\frac{1}{\sqrt{\omega_k^2+s}}\left(\frac{1}{e^{\frac{\sqrt{\omega_k^2+s}}{T}}-1}+\frac{1}{2}\right)\right).
\end{eqnarray}
Here we have taken care of the minus sign due to the clockwise contours.
Integrating it back and setting $s=0$, we arrive at (apart from a constant) 
\begin{equation}
I(0)=\frac{1}{2}\int d^4x \int \frac{d^3k}{\left(2\pi\right)^3}
\left(T \ln \left(1-e^{-\frac{\omega_k }{T}}\right)+\frac{1}{2} \omega_k\right).
\end{equation}

\subsection{Renormalization of $V_{\text{eff}}$}\label{app:Renormalization}
Here we follow Ref.~\cite{ChienPRA12} to renormalize $V_{eff}$ of leading-order large-$\mathcal{N}$ theory. The renormalized coupling constant can be defined as
\begin{equation}
\frac{1}{\lambda _R}=-\frac{\delta  V_{\text{eff}}}{\delta \alpha \delta \alpha}.
\end{equation}
For the leading-order large-$\mathcal{N}$ theory, $\frac{\delta  V_{\text{eff}}}{\delta \alpha \delta \alpha     }=-\frac{1}{\lambda }+$ (regular terms). Thus, the renormalized coupling constant can be identified as the bare coupling constant. This lead to $\lambda_R=(4\pi\hbar^2 a/m)$, where $a$ is the two-body $s$-wave scattering length \cite{pethick2008bose}. By inspecting the classical part $-\frac{\partial V_{\text{eff}}}{\partial \chi }|_{\chi=0}=\frac{\mu }{\lambda }$,  the renormalized chemical potential $\mu_R$ is given by 
\begin{equation}
-\frac{\partial V_{\text{eff}}}{\partial \chi }=\frac{\mu _R}{\lambda }=\frac{\mu }{\lambda }-\sum _k \frac{1}{2}.
\end{equation}

The renormalized effective potential is 
\begin{equation}
V_{R,\text{eff}}=V_{R,0}+\sum _k \frac{\ln \left(1-e^{-\beta  \omega _k}\right)}{\beta }-\frac{\left(\mu _R+\chi \right){}^2}{2 \lambda }+\chi \phi ^* \phi,
\end{equation}
where $V_{R,0}$ is an infinite constant that absorbs the zero-point energy and may be dropped. We define $\alpha_R=\mu_R+\chi$, too.

\subsection{Derivation of the LOAF theory}\label{app:LOAF}
Here we briefly review the LOAF theory by skipping the derivation with $\mathcal{N}$ copies of the fields and just presenting the leading-order $1/\mathcal{N}$ theory with $\mathcal{N}$ set to $1$. The detailed derivation can be found in Refs.~\cite{PhysRevLett.105.240402,PhysRevA.83.053622}. In the LOAF theory, we introduce two auxiliary fields $\chi$ and $A$ representing the normal and anomalous densities. They can be introduced by inserting the following identity into the partition function~\eqref{eq:partition}. 
\begin{eqnarray}
1
&&=
\int\mathcal{D}\chi  \text{$\mathcal{D}$A} \text{$\mathcal{D}$A}^* \nonumber\\
&&\delta  \left(\chi -\sqrt{2} \lambda  \phi ^* \phi \right) \delta  \left(A-\lambda  \phi ^2\right) \delta  \left(A^*-\lambda  \left(\phi ^*\right)^2\right)\nonumber\\
&&=\mathcal{C}^{\prime}\int\mathcal{D}\chi  \mathcal{D} \tilde{\chi } \text{$\mathcal{D}$A} \mathcal{D} \tilde{A} \text{$\mathcal{D}$A}^* \mathcal{D} \tilde{A^*}\nonumber\\
&& e^{\frac{\tilde{\chi } \left(\chi -\sqrt{2} \lambda  \phi ^* \phi \right)}{\lambda }} e^{\frac{\tilde{A^*} \left(A-\lambda  \phi ^2\right)}{\lambda }} e^{\frac{\tilde{A} \left(A^*-\lambda  \left(\phi ^*\right)^2\right)}{\lambda }}.
\end{eqnarray}
Here $\mathcal{C}^{\prime}$ is a normalization factor, and the contour of integration follows the description below Eq.~\eqref{eq:dirac_delta}. 
Then the quartic term in $\phi$ can be replaced by 
\begin{eqnarray}
\frac{1}{2} \lambda  \left(\phi ^* \phi \right)^2
&&=\frac{1}{2} \left(2 \lambda  \left(\phi ^* \phi \right)^2-\lambda  \left(\phi ^*\right)^2 \phi ^2\right)\nonumber\\
&&=\frac{1}{2} \left(2 \lambda  \left(\frac{\chi }{\sqrt{2} \lambda }\right)^2-\frac{\lambda  A A^*}{\lambda  \lambda }\right).
\end{eqnarray}
With the quartic terms replaced, the action becomes
\begin{eqnarray}
&&\int[dx](\frac{1}{2} \Phi \bar{G}_0^{-1} \Phi-\frac{\mu  \chi }{\sqrt{2} \lambda }+\frac{\chi ^2}{2 \lambda }-\frac{A A^*}{2 \lambda }-J^{\dagger} \Phi\nonumber\\
&& -\frac{\tilde{\chi } \chi }{\lambda }-\frac{\tilde{A^*} A}{\lambda }-\frac{\tilde{A} A^*}{\lambda }
-s\chi-\mathcal{S}^*A-\mathcal{S} A^*),
\end{eqnarray}
where $s$, $\mathcal{S}$, and $\mathcal{S}^*$ are the source terms for the auxiliary fields, and 
$\bar{G}_0^{-1}=\tilde{G}_0^{-1}+\left(
\begin{array}{cc}
 \sqrt{2} \tilde{\chi } & 2 \tilde{A^*} \\
 2 \tilde{A} & \sqrt{2} \tilde{\chi } \\
\end{array}
\right)$. 
Here $\tilde{G}_0^{-1}=diag(\frac{\partial }{\partial \tau } -\frac{\nabla^2}{2m} ,-\frac{\partial }{\partial \tau }- \frac{\nabla^2}{2m})$.
Performing the $\phi$ integral, 
the effective action becomes
\begin{eqnarray}
S_{\text{eff}}
&&=
\frac{1}{2} \text{Tr}\left[\ln \left(\bar{G}_0^{-1}\right)\right]
+\int[dx]\left(-\frac{1}{2} J^{\dagger} \bar{G}_0 J-\frac{\mu  \chi }{\sqrt{2} \lambda }+\frac{\chi ^2}{2 \lambda }-\right.
\nonumber\\
&&\left.\frac{A A^*}{2 \lambda }-\frac{\tilde{\chi } \chi }{\lambda }-\frac{\tilde{A^*} A}{\lambda }-\frac{\tilde{A} A^*}{\lambda }-s\chi-\mathcal{S}^*A-\mathcal{S} A^*\right).
\end{eqnarray}
To obtain the effective potential, we apply a Legendre transform and replace $J$ in terms of $\phi_c$ and $\phi^*_c$ by using $J=\int[dx]\bar{G}_0^{-1} \Phi_c$. This leads to the grand potential 
\begin{eqnarray}
\Gamma&&=\int[dx](\Phi_c  J^{\dagger}+s\chi_c+\mathcal{S}^*A_c+\mathcal{S} A_c^*)+S_{\text{eff}}\nonumber\\
=&&
\frac{1}{2} \text{Tr}\left[\ln \left(\bar{G}_0^{-1}\right)\right]
+\int[dx]\left(\Phi ^{\dagger} \bar{G}_0^{-1} \Phi -\frac{\mu  \chi }{\sqrt{2} \lambda }+\frac{\chi ^2}{2 \lambda }-\right.\nonumber\\
&&\left.\frac{A A^*}{2 \lambda }-\frac{\tilde{\chi } \chi }{\lambda }-\frac{\tilde{A^*} A}{\lambda }-\frac{\tilde{A} A^*}{\lambda }\right).
\end{eqnarray}
In the following we drop the subscript $c$ denoting the expectation values.

The equilibrium state corresponds to the theory at the minimum of $\Gamma$. Thus, 
\begin{eqnarray}
\frac{\delta \Gamma }{\delta \chi }
=0\Rightarrow \tilde{\chi }
=\chi -\frac{\mu }{\sqrt{2}}.\\
\frac{\delta \Gamma }{\text{$\delta $A}^*}
=0\Rightarrow \tilde{A}=-\frac{A}{2}.\\
\frac{\delta \Gamma }{\text{$\delta $A}}
=0\Rightarrow \tilde{A^*}=-\frac{A^*}{2}.
\end{eqnarray}
At the minimum, $\Gamma$ has the expression 
\begin{equation}
\Gamma
=\int[dx]( \Phi ^{\dagger} \bar{G}_0^{-1} \Phi -\frac{\chi ^2}{2 \lambda }+\frac{A A^*}{2 \lambda })+\frac{1}{2} \text{Tr}\left[\ln \left(\bar{G}_0^{-1}\right)\right],	
\end{equation}
where 
\begin{eqnarray}
\bar{G}_0^{-1}
&&= \left(
\begin{array}{cc}
 \partial _{\tau }-\frac{\nabla ^2}{2 m}+\chi^{\prime} & -A^* \\
 -A & -\partial _{\tau }-\frac{\nabla ^2}{2 m}+\chi^{\prime} \\
\end{array}
\right).
\end{eqnarray}
Here $\chi^{\prime}=\sqrt{2} \chi -\mu$.
Following Appendix~\ref{app:TrLn}, for homogeneous static fields we obtain 
\begin{equation}
\frac{1}{2} \text{Tr}\left[\ln \left(\bar{G}_0^{-1}\right)\right]
=\int[dx]\sum_{k}\left[T \ln \left(1-e^{-\frac{\omega _k}{T}}\right)+\frac{\omega _k}{2}\right],
\end{equation}
where $\omega _k=\sqrt{\left(\epsilon _k+\chi^{\prime}\right){}^2-A A^*}$ and the dispersion is gapless in the presence of BEC. For homogeneous and static fields, the effective potential $V_{\text{eff}}=\Gamma/(\beta\Omega)$ becomes 
\begin{eqnarray}
V_{\text{eff}}
&&=
 \chi^{\prime}  \phi ^* \phi  -\frac{1}{2} A^* \left(\phi ^*\right)^2-\frac{A \phi ^2}{2}-\frac{\left(\chi^{\prime}+\mu\right) ^2}{4 \lambda }+\frac{A A^*}{2 \lambda }+\nonumber\\
&&\sum _k \left(\frac{\omega _k}{2}+T \ln \left(1-e^{-\frac{\omega _k}{T}}\right)\right).
\end{eqnarray}
The ultraviolet divergence of $V_{eff}$ can be renormalized in the normal and BEC phases, while in the intermediate superfluid phase it can be regularized. The regularization smoothly interpolates the two renormalizations at high and low temperatures. Following the procedures described in Refs.~\cite{PhysRevLett.105.240402,PhysRevA.83.053622}, the effective potential after the renormalization and regularization is shown in Eq.~\eqref{eq:2fVReff}. Then minimizing $V_{eff}$ with respect to the fields and implementing standard thermodynamic relations lead to the equations of state shown in Eqs.~\eqref{eq:2fmu}-\eqref{eq:2fA}.

\bibliographystyle{apsrev4-1}

\begin{thebibliography}{37}%
	\makeatletter
	\providecommand \@ifxundefined [1]{%
		\@ifx{#1\undefined}
	}%
	\providecommand \@ifnum [1]{%
		\ifnum #1\expandafter \@firstoftwo
		\else \expandafter \@secondoftwo
		\fi
	}%
	\providecommand \@ifx [1]{%
		\ifx #1\expandafter \@firstoftwo
		\else \expandafter \@secondoftwo
		\fi
	}%
	\providecommand \natexlab [1]{#1}%
	\providecommand \enquote  [1]{``#1''}%
	\providecommand \bibnamefont  [1]{#1}%
	\providecommand \bibfnamefont [1]{#1}%
	\providecommand \citenamefont [1]{#1}%
	\providecommand \href@noop [0]{\@secondoftwo}%
	\providecommand \href [0]{\begingroup \@sanitize@url \@href}%
	\providecommand \@href[1]{\@@startlink{#1}\@@href}%
	\providecommand \@@href[1]{\endgroup#1\@@endlink}%
	\providecommand \@sanitize@url [0]{\catcode `\\12\catcode `\$12\catcode
		`\&12\catcode `\#12\catcode `\^12\catcode `\_12\catcode `\%12\relax}%
	\providecommand \@@startlink[1]{}%
	\providecommand \@@endlink[0]{}%
	\providecommand \url  [0]{\begingroup\@sanitize@url \@url }%
	\providecommand \@url [1]{\endgroup\@href {#1}{\urlprefix }}%
	\providecommand \urlprefix  [0]{URL }%
	\providecommand \Eprint [0]{\href }%
	\providecommand \doibase [0]{http://dx.doi.org/}%
	\providecommand \selectlanguage [0]{\@gobble}%
	\providecommand \bibinfo  [0]{\@secondoftwo}%
	\providecommand \bibfield  [0]{\@secondoftwo}%
	\providecommand \translation [1]{[#1]}%
	\providecommand \BibitemOpen [0]{}%
	\providecommand \bibitemStop [0]{}%
	\providecommand \bibitemNoStop [0]{.\EOS\space}%
	\providecommand \EOS [0]{\spacefactor3000\relax}%
	\providecommand \BibitemShut  [1]{\csname bibitem#1\endcsname}%
	\let\auto@bib@innerbib\@empty
	\bibitem [{\citenamefont {Pethick}\ and\ \citenamefont
		{Smith}(2008)}]{pethick2008bose}%
	\BibitemOpen
	\bibfield  {author} {\bibinfo {author} {\bibfnamefont {C.}~\bibnamefont
			{Pethick}}\ and\ \bibinfo {author} {\bibfnamefont {H.}~\bibnamefont
			{Smith}},\ }\href {https://books.google.com/books?id=G8kgAwAAQBAJ} {\emph
		{\bibinfo {title} {Bose--Einstein Condensation in Dilute Gases}}}\ (\bibinfo
	{publisher} {Cambridge University Press},\ \bibinfo {year}
	{2008})\BibitemShut {NoStop}%
	\bibitem [{\citenamefont {Ueda}(2010)}]{ueda2010fundamentals}%
	\BibitemOpen
	\bibfield  {author} {\bibinfo {author} {\bibfnamefont {M.}~\bibnamefont
			{Ueda}},\ }\href {https://books.google.com/books?id=iix3\_pqy6ysC} {\emph
		{\bibinfo {title} {Fundamentals and New Frontiers of Bose-Einstein
				Condensation}}}\ (\bibinfo  {publisher} {World Scientific},\ \bibinfo {year}
	{2010})\BibitemShut {NoStop}%
	\bibitem [{\citenamefont {Stoof}\ \emph {et~al.}(2008)\citenamefont {Stoof},
		\citenamefont {Gubbels},\ and\ \citenamefont
		{Dickerscheid}}]{stoof2008ultracold}%
	\BibitemOpen
	\bibfield  {author} {\bibinfo {author} {\bibfnamefont {H.}~\bibnamefont
			{Stoof}}, \bibinfo {author} {\bibfnamefont {K.}~\bibnamefont {Gubbels}}, \
		and\ \bibinfo {author} {\bibfnamefont {D.}~\bibnamefont {Dickerscheid}},\
	}\href {https://books.google.com/books?id=-9c2Ns2J5P4C} {\emph {\bibinfo
		{title} {Ultracold Quantum Fields}}},\ Theoretical and Mathematical Physics\
(\bibinfo  {publisher} {Springer},\ \bibinfo {year} {2008})\BibitemShut
{NoStop}%
\bibitem [{\citenamefont {Andersen}(2004)}]{RevModPhys.76.599}%
\BibitemOpen
\bibfield  {author} {\bibinfo {author} {\bibfnamefont {J.~O.}\ \bibnamefont
		{Andersen}},\ }\href {\doibase 10.1103/RevModPhys.76.599} {\bibfield
	{journal} {\bibinfo  {journal} {Rev. Mod. Phys.}\ }\textbf {\bibinfo {volume}
		{76}},\ \bibinfo {pages} {599} (\bibinfo {year} {2004})}\BibitemShut
{NoStop}%
\bibitem [{\citenamefont {Dalfovo}\ \emph {et~al.}(1999)\citenamefont
	{Dalfovo}, \citenamefont {Giorgini}, \citenamefont {Pitaevskii},\ and\
	\citenamefont {Stringari}}]{RevModPhys.71.463}%
\BibitemOpen
\bibfield  {author} {\bibinfo {author} {\bibfnamefont {F.}~\bibnamefont
		{Dalfovo}}, \bibinfo {author} {\bibfnamefont {S.}~\bibnamefont {Giorgini}},
	\bibinfo {author} {\bibfnamefont {L.~P.}\ \bibnamefont {Pitaevskii}}, \ and\
	\bibinfo {author} {\bibfnamefont {S.}~\bibnamefont {Stringari}},\ }\href
{\doibase 10.1103/RevModPhys.71.463} {\bibfield  {journal} {\bibinfo
		{journal} {Rev. Mod. Phys.}\ }\textbf {\bibinfo {volume} {71}},\ \bibinfo
	{pages} {463} (\bibinfo {year} {1999})}\BibitemShut {NoStop}%
\bibitem [{\citenamefont {Bloch}\ \emph {et~al.}(2008)\citenamefont {Bloch},
	\citenamefont {Dalibard},\ and\ \citenamefont {Zwerger}}]{RevModPhys.80.885}%
\BibitemOpen
\bibfield  {author} {\bibinfo {author} {\bibfnamefont {I.}~\bibnamefont
		{Bloch}}, \bibinfo {author} {\bibfnamefont {J.}~\bibnamefont {Dalibard}}, \
	and\ \bibinfo {author} {\bibfnamefont {W.}~\bibnamefont {Zwerger}},\ }\href
{\doibase 10.1103/RevModPhys.80.885} {\bibfield  {journal} {\bibinfo
		{journal} {Rev. Mod. Phys.}\ }\textbf {\bibinfo {volume} {80}},\ \bibinfo
	{pages} {885} (\bibinfo {year} {2008})}\BibitemShut {NoStop}%
\bibitem [{\citenamefont {Giorgini}\ \emph {et~al.}(1997)\citenamefont
	{Giorgini}, \citenamefont {Pitaevskii},\ and\ \citenamefont
	{Stringari}}]{BF02396737}%
\BibitemOpen
\bibfield  {author} {\bibinfo {author} {\bibfnamefont {S.}~\bibnamefont
		{Giorgini}}, \bibinfo {author} {\bibfnamefont {L.}~\bibnamefont
		{Pitaevskii}}, \ and\ \bibinfo {author} {\bibfnamefont {S.}~\bibnamefont
		{Stringari}},\ }\href {\doibase 10.1007/BF02396737} {\bibfield  {journal}
	{\bibinfo  {journal} {Journal of Low Temperature Physics}\ }\textbf {\bibinfo
		{volume} {109}},\ \bibinfo {pages} {309} (\bibinfo {year}
	{1997})}\BibitemShut {NoStop}%
\bibitem [{\citenamefont {Gerbier}\ \emph {et~al.}(2004)\citenamefont
	{Gerbier}, \citenamefont {Thywissen}, \citenamefont {Richard}, \citenamefont
	{Hugbart}, \citenamefont {Bouyer},\ and\ \citenamefont
	{Aspect}}]{PhysRevLett.92.030405}%
\BibitemOpen
\bibfield  {author} {\bibinfo {author} {\bibfnamefont {F.}~\bibnamefont
		{Gerbier}}, \bibinfo {author} {\bibfnamefont {J.~H.}\ \bibnamefont
		{Thywissen}}, \bibinfo {author} {\bibfnamefont {S.}~\bibnamefont {Richard}},
	\bibinfo {author} {\bibfnamefont {M.}~\bibnamefont {Hugbart}}, \bibinfo
	{author} {\bibfnamefont {P.}~\bibnamefont {Bouyer}}, \ and\ \bibinfo {author}
	{\bibfnamefont {A.}~\bibnamefont {Aspect}},\ }\href {\doibase
	10.1103/PhysRevLett.92.030405} {\bibfield  {journal} {\bibinfo  {journal}
		{Phys. Rev. Lett.}\ }\textbf {\bibinfo {volume} {92}},\ \bibinfo {pages}
	{030405} (\bibinfo {year} {2004})}\BibitemShut {NoStop}%
\bibitem [{\citenamefont {Meppelink}\ \emph {et~al.}(2010)\citenamefont
	{Meppelink}, \citenamefont {Rozendaal}, \citenamefont {Koller}, \citenamefont
	{Vogels},\ and\ \citenamefont {van~der Straten}}]{PhysRevA.81.053632}%
\BibitemOpen
\bibfield  {author} {\bibinfo {author} {\bibfnamefont {R.}~\bibnamefont
		{Meppelink}}, \bibinfo {author} {\bibfnamefont {R.~A.}\ \bibnamefont
		{Rozendaal}}, \bibinfo {author} {\bibfnamefont {S.~B.}\ \bibnamefont
		{Koller}}, \bibinfo {author} {\bibfnamefont {J.~M.}\ \bibnamefont {Vogels}},
	\ and\ \bibinfo {author} {\bibfnamefont {P.}~\bibnamefont {van~der
			Straten}},\ }\href {\doibase 10.1103/PhysRevA.81.053632} {\bibfield
	{journal} {\bibinfo  {journal} {Phys. Rev. A}\ }\textbf {\bibinfo {volume}
		{81}},\ \bibinfo {pages} {053632} (\bibinfo {year} {2010})}\BibitemShut
{NoStop}%
\bibitem [{\citenamefont {Smith}\ \emph {et~al.}(2011)\citenamefont {Smith},
	\citenamefont {Campbell}, \citenamefont {Tammuz},\ and\ \citenamefont
	{Hadzibabic}}]{PhysRevLett.106.250403}%
\BibitemOpen
\bibfield  {author} {\bibinfo {author} {\bibfnamefont {R.~P.}\ \bibnamefont
		{Smith}}, \bibinfo {author} {\bibfnamefont {R.~L.~D.}\ \bibnamefont
		{Campbell}}, \bibinfo {author} {\bibfnamefont {N.}~\bibnamefont {Tammuz}}, \
	and\ \bibinfo {author} {\bibfnamefont {Z.}~\bibnamefont {Hadzibabic}},\
}\href {\doibase 10.1103/PhysRevLett.106.250403} {\bibfield  {journal}
{\bibinfo  {journal} {Phys. Rev. Lett.}\ }\textbf {\bibinfo {volume} {106}},\
\bibinfo {pages} {250403} (\bibinfo {year} {2011})}\BibitemShut {NoStop}%
\bibitem [{\citenamefont {Giorgini}\ \emph {et~al.}(1996)\citenamefont
	{Giorgini}, \citenamefont {Pitaevskii},\ and\ \citenamefont
	{Stringari}}]{PhysRevA.54.R4633}%
\BibitemOpen
\bibfield  {author} {\bibinfo {author} {\bibfnamefont {S.}~\bibnamefont
		{Giorgini}}, \bibinfo {author} {\bibfnamefont {L.~P.}\ \bibnamefont
		{Pitaevskii}}, \ and\ \bibinfo {author} {\bibfnamefont {S.}~\bibnamefont
		{Stringari}},\ }\href {\doibase 10.1103/PhysRevA.54.R4633} {\bibfield
	{journal} {\bibinfo  {journal} {Phys. Rev. A}\ }\textbf {\bibinfo {volume}
		{54}},\ \bibinfo {pages} {R4633} (\bibinfo {year} {1996})}\BibitemShut
{NoStop}%
\bibitem [{\citenamefont {Cooper}\ \emph {et~al.}(2010)\citenamefont {Cooper},
	\citenamefont {Chien}, \citenamefont {Mihaila}, \citenamefont {Dawson},\ and\
	\citenamefont {Timmermans}}]{PhysRevLett.105.240402}%
\BibitemOpen
\bibfield  {author} {\bibinfo {author} {\bibfnamefont {F.}~\bibnamefont
		{Cooper}}, \bibinfo {author} {\bibfnamefont {C.-C.}\ \bibnamefont {Chien}},
	\bibinfo {author} {\bibfnamefont {B.}~\bibnamefont {Mihaila}}, \bibinfo
	{author} {\bibfnamefont {J.~F.}\ \bibnamefont {Dawson}}, \ and\ \bibinfo
	{author} {\bibfnamefont {E.}~\bibnamefont {Timmermans}},\ }\href {\doibase
	10.1103/PhysRevLett.105.240402} {\bibfield  {journal} {\bibinfo  {journal}
		{Phys. Rev. Lett.}\ }\textbf {\bibinfo {volume} {105}},\ \bibinfo {pages}
	{240402} (\bibinfo {year} {2010})}\BibitemShut {NoStop}%
\bibitem [{\citenamefont {Cooper}\ \emph {et~al.}(2012)\citenamefont {Cooper},
	\citenamefont {Chien}, \citenamefont {Mihaila}, \citenamefont {Dawson},\ and\
	\citenamefont {Timmermans}}]{PhysRevA.85.023631}%
\BibitemOpen
\bibfield  {author} {\bibinfo {author} {\bibfnamefont {F.}~\bibnamefont
		{Cooper}}, \bibinfo {author} {\bibfnamefont {C.-C.}\ \bibnamefont {Chien}},
	\bibinfo {author} {\bibfnamefont {B.}~\bibnamefont {Mihaila}}, \bibinfo
	{author} {\bibfnamefont {J.~F.}\ \bibnamefont {Dawson}}, \ and\ \bibinfo
	{author} {\bibfnamefont {E.}~\bibnamefont {Timmermans}},\ }\href {\doibase
	10.1103/PhysRevA.85.023631} {\bibfield  {journal} {\bibinfo  {journal} {Phys.
			Rev. A}\ }\textbf {\bibinfo {volume} {85}},\ \bibinfo {pages} {023631}
	(\bibinfo {year} {2012})}\BibitemShut {NoStop}%
\bibitem [{\citenamefont {Chien}\ \emph {et~al.}(2012)\citenamefont {Chien},
	\citenamefont {Cooper},\ and\ \citenamefont {Timmermans}}]{ChienPRA12}%
\BibitemOpen
\bibfield  {author} {\bibinfo {author} {\bibfnamefont {C.-C.}\ \bibnamefont
		{Chien}}, \bibinfo {author} {\bibfnamefont {F.}~\bibnamefont {Cooper}}, \
	and\ \bibinfo {author} {\bibfnamefont {E.}~\bibnamefont {Timmermans}},\
}\href {\doibase 10.1103/PhysRevA.86.023634} {\bibfield  {journal} {\bibinfo
	{journal} {Phys. Rev. A}\ }\textbf {\bibinfo {volume} {86}},\ \bibinfo
{pages} {023634} (\bibinfo {year} {2012})}\BibitemShut {NoStop}%
\bibitem [{\citenamefont {Chien}\ \emph {et~al.}(2014)\citenamefont {Chien},
	\citenamefont {She},\ and\ \citenamefont {Cooper}}]{ChienAnnPhys}%
\BibitemOpen
\bibfield  {author} {\bibinfo {author} {\bibfnamefont {C.~C.}\ \bibnamefont
		{Chien}}, \bibinfo {author} {\bibfnamefont {J.~H.}\ \bibnamefont {She}}, \
	and\ \bibinfo {author} {\bibfnamefont {F.}~\bibnamefont {Cooper}},\ }\href
{http://www.sciencedirect.com/science/article/pii/S0003491614001006}
{\bibfield  {journal} {\bibinfo  {journal} {Ann. Phys.}\ }\textbf {\bibinfo
		{volume} {347}} (\bibinfo {year} {2014})}\BibitemShut {NoStop}%
\bibitem [{\citenamefont {Liu}\ \emph {et~al.}(2015)\citenamefont {Liu},
	\citenamefont {Mulkerin}, \citenamefont {He},\ and\ \citenamefont
	{Hu}}]{PhysRevA.91.043631}%
\BibitemOpen
\bibfield  {author} {\bibinfo {author} {\bibfnamefont {X.~J.}\ \bibnamefont
		{Liu}}, \bibinfo {author} {\bibfnamefont {B.}~\bibnamefont {Mulkerin}},
	\bibinfo {author} {\bibfnamefont {L.}~\bibnamefont {He}}, \ and\ \bibinfo
	{author} {\bibfnamefont {H.}~\bibnamefont {Hu}},\ }\href
{http://journals.aps.org/pra/abstract/10.1103/PhysRevA.91.043631} {\bibfield
	{journal} {\bibinfo  {journal} {Phys. Rev. A}\ }\textbf {\bibinfo {volume}
		{91}},\ \bibinfo {pages} {043631} (\bibinfo {year} {2015})}\BibitemShut
{NoStop}%
\bibitem [{\citenamefont {Chien}\ \emph {et~al.}(2007)\citenamefont {Chien},
	\citenamefont {Chen}, \citenamefont {He},\ and\ \citenamefont
	{Levin}}]{PhysRevLett.98.110404}%
\BibitemOpen
\bibfield  {author} {\bibinfo {author} {\bibfnamefont {C.-C.}\ \bibnamefont
		{Chien}}, \bibinfo {author} {\bibfnamefont {Q.}~\bibnamefont {Chen}},
	\bibinfo {author} {\bibfnamefont {Y.}~\bibnamefont {He}}, \ and\ \bibinfo
	{author} {\bibfnamefont {K.}~\bibnamefont {Levin}},\ }\href {\doibase
	10.1103/PhysRevLett.98.110404} {\bibfield  {journal} {\bibinfo  {journal}
		{Phys. Rev. Lett.}\ }\textbf {\bibinfo {volume} {98}},\ \bibinfo {pages}
	{110404} (\bibinfo {year} {2007})}\BibitemShut {NoStop}%
\bibitem [{\citenamefont {Stewart}\ \emph {et~al.}(2008)\citenamefont
	{Stewart}, \citenamefont {Gaebler},\ and\ \citenamefont {Jin}}]{Stewart2008}%
\BibitemOpen
\bibfield  {author} {\bibinfo {author} {\bibfnamefont {J.~T.}\ \bibnamefont
		{Stewart}}, \bibinfo {author} {\bibfnamefont {J.~P.}\ \bibnamefont
		{Gaebler}}, \ and\ \bibinfo {author} {\bibfnamefont {D.~S.}\ \bibnamefont
		{Jin}},\ }\href {\doibase 10.1038/nature07172} {\bibfield  {journal}
	{\bibinfo  {journal} {Nature}\ }\textbf {\bibinfo {volume} {454}},\ \bibinfo
	{pages} {744} (\bibinfo {year} {2008})}\BibitemShut {NoStop}%
\bibitem [{\citenamefont {Fetter}\ and\ \citenamefont
	{Walecka}(2012)}]{fetter2012quantum}%
\BibitemOpen
\bibfield  {author} {\bibinfo {author} {\bibfnamefont {A.}~\bibnamefont
		{Fetter}}\ and\ \bibinfo {author} {\bibfnamefont {J.}~\bibnamefont
		{Walecka}},\ }\href {https://books.google.com/books?id=t5\_DAgAAQBAJ} {\emph
	{\bibinfo {title} {Quantum Theory of Many-Particle Systems}}},\ Dover Books
on Physics\ (\bibinfo  {publisher} {Dover Publications},\ \bibinfo {year}
{2012})\BibitemShut {NoStop}%
\bibitem [{\citenamefont {Haugerud}\ \emph {et~al.}(1997)\citenamefont
	{Haugerud}, \citenamefont {Haugset},\ and\ \citenamefont
	{Ravndal}}]{Haugerud199718}%
\BibitemOpen
\bibfield  {author} {\bibinfo {author} {\bibfnamefont {H.}~\bibnamefont
		{Haugerud}}, \bibinfo {author} {\bibfnamefont {T.}~\bibnamefont {Haugset}}, \
	and\ \bibinfo {author} {\bibfnamefont {F.}~\bibnamefont {Ravndal}},\ }\href
{\doibase 10.1016/S0375-9601(96)08842-1} {\bibfield  {journal} {\bibinfo
		{journal} {Physics Letters A}\ }\textbf {\bibinfo {volume} {225}},\ \bibinfo
	{pages} {18 } (\bibinfo {year} {1997})}\BibitemShut {NoStop}%
\bibitem [{\citenamefont {Grossmann}\ and\ \citenamefont
	{Holthaus}(1995)}]{Grossmann1995188}%
\BibitemOpen
\bibfield  {author} {\bibinfo {author} {\bibfnamefont {S.}~\bibnamefont
		{Grossmann}}\ and\ \bibinfo {author} {\bibfnamefont {M.}~\bibnamefont
		{Holthaus}},\ }\href {\doibase 10.1016/0375-9601(95)00766-V} {\bibfield
	{journal} {\bibinfo  {journal} {Physics Letters A}\ }\textbf {\bibinfo
		{volume} {208}},\ \bibinfo {pages} {188 } (\bibinfo {year}
	{1995})}\BibitemShut {NoStop}%
\bibitem [{\citenamefont {Mazenko}(2000)}]{mazenko2000equilibrium}%
\BibitemOpen
\bibfield  {author} {\bibinfo {author} {\bibfnamefont {G.}~\bibnamefont
		{Mazenko}},\ }\href {https://books.google.com/books?id=Vn5GAAAAYAAJ} {\emph
	{\bibinfo {title} {Equilibrium statistical mechanics}}},\ \bibinfo {series}
{A Wiley-interscience publication}\ No.\ \bibinfo {number} {v. 1}\ (\bibinfo
{publisher} {Wiley},\ \bibinfo {year} {2000})\BibitemShut {NoStop}%
\bibitem [{\citenamefont {Ketterle}\ and\ \citenamefont {van
		Druten}(1996)}]{PhysRevA.54.656}%
\BibitemOpen
\bibfield  {author} {\bibinfo {author} {\bibfnamefont {W.}~\bibnamefont
		{Ketterle}}\ and\ \bibinfo {author} {\bibfnamefont {N.~J.}\ \bibnamefont {van
			Druten}},\ }\href {\doibase 10.1103/PhysRevA.54.656} {\bibfield  {journal}
	{\bibinfo  {journal} {Phys. Rev. A}\ }\textbf {\bibinfo {volume} {54}},\
	\bibinfo {pages} {656} (\bibinfo {year} {1996})}\BibitemShut {NoStop}%
\bibitem [{\citenamefont {Kirsten}\ and\ \citenamefont
	{Toms}(1996)}]{PhysRevA.54.4188}%
\BibitemOpen
\bibfield  {author} {\bibinfo {author} {\bibfnamefont {K.}~\bibnamefont
		{Kirsten}}\ and\ \bibinfo {author} {\bibfnamefont {D.~J.}\ \bibnamefont
		{Toms}},\ }\href {\doibase 10.1103/PhysRevA.54.4188} {\bibfield  {journal}
	{\bibinfo  {journal} {Phys. Rev. A}\ }\textbf {\bibinfo {volume} {54}},\
	\bibinfo {pages} {4188} (\bibinfo {year} {1996})}\BibitemShut {NoStop}%
\bibitem [{\citenamefont {Pathria}(1998)}]{PhysRevA.58.1490}%
\BibitemOpen
\bibfield  {author} {\bibinfo {author} {\bibfnamefont {R.~K.}\ \bibnamefont
		{Pathria}},\ }\href {\doibase 10.1103/PhysRevA.58.1490} {\bibfield  {journal}
	{\bibinfo  {journal} {Phys. Rev. A}\ }\textbf {\bibinfo {volume} {58}},\
	\bibinfo {pages} {1490} (\bibinfo {year} {1998})}\BibitemShut {NoStop}%
\bibitem [{\citenamefont {Jaouadi}\ \emph {et~al.}(2011)\citenamefont
	{Jaouadi}, \citenamefont {Telmini},\ and\ \citenamefont
	{Charron}}]{PhysRevA.83.023616}%
\BibitemOpen
\bibfield  {author} {\bibinfo {author} {\bibfnamefont {A.}~\bibnamefont
		{Jaouadi}}, \bibinfo {author} {\bibfnamefont {M.}~\bibnamefont {Telmini}}, \
	and\ \bibinfo {author} {\bibfnamefont {E.}~\bibnamefont {Charron}},\ }\href
{\doibase 10.1103/PhysRevA.83.023616} {\bibfield  {journal} {\bibinfo
		{journal} {Phys. Rev. A}\ }\textbf {\bibinfo {volume} {83}},\ \bibinfo
	{pages} {023616} (\bibinfo {year} {2011})}\BibitemShut {NoStop}%
\bibitem [{\citenamefont {Noronha}(2015)}]{PhysRevA.92.017601}%
\BibitemOpen
\bibfield  {author} {\bibinfo {author} {\bibfnamefont {J.~M.~B.}\
		\bibnamefont {Noronha}},\ }\href {\doibase 10.1103/PhysRevA.92.017601}
{\bibfield  {journal} {\bibinfo  {journal} {Phys. Rev. A}\ }\textbf {\bibinfo
		{volume} {92}},\ \bibinfo {pages} {017601} (\bibinfo {year}
	{2015})}\BibitemShut {NoStop}%
\bibitem [{\citenamefont {Arnold}\ and\ \citenamefont
	{Tom\'a\ifmmode~\check{s}\else \v{s}\fi{}ik}(2001)}]{PhysRevA.64.053609}%
\BibitemOpen
\bibfield  {author} {\bibinfo {author} {\bibfnamefont {P.}~\bibnamefont
		{Arnold}}\ and\ \bibinfo {author} {\bibfnamefont {B.}~\bibnamefont
		{Tom\'a\ifmmode~\check{s}\else \v{s}\fi{}ik}},\ }\href {\doibase
	10.1103/PhysRevA.64.053609} {\bibfield  {journal} {\bibinfo  {journal} {Phys.
			Rev. A}\ }\textbf {\bibinfo {volume} {64}},\ \bibinfo {pages} {053609}
	(\bibinfo {year} {2001})}\BibitemShut {NoStop}%
\bibitem [{\citenamefont {Popov}\ and\ \citenamefont
	{Popov}(1991)}]{popov1991functional}%
\BibitemOpen
\bibfield  {author} {\bibinfo {author} {\bibfnamefont {V.}~\bibnamefont
		{Popov}}\ and\ \bibinfo {author} {\bibfnamefont {V.}~\bibnamefont {Popov}},\
}\href {https://books.google.com/books?id=Bx59iYMgB84C} {\emph {\bibinfo
	{title} {Functional Integrals and Collective Excitations}}},\ Cambridge
Monographs on Mathematical Physics\ (\bibinfo  {publisher} {Cambridge
	University Press},\ \bibinfo {year} {1991})\BibitemShut {NoStop}%
\bibitem [{\citenamefont {Baym}\ \emph {et~al.}(2000)\citenamefont {Baym},
	\citenamefont {Blaizot},\ and\ \citenamefont
	{Zinn-Justin}}]{0295-5075-49-2-150}%
\BibitemOpen
\bibfield  {author} {\bibinfo {author} {\bibfnamefont {G.}~\bibnamefont
		{Baym}}, \bibinfo {author} {\bibfnamefont {J.-P.}\ \bibnamefont {Blaizot}}, \
	and\ \bibinfo {author} {\bibfnamefont {J.}~\bibnamefont {Zinn-Justin}},\
}\href {http://stacks.iop.org/0295-5075/49/i=2/a=150} {\bibfield  {journal}
{\bibinfo  {journal} {EPL (Europhysics Letters)}\ }\textbf {\bibinfo {volume}
	{49}},\ \bibinfo {pages} {150} (\bibinfo {year} {2000})}\BibitemShut
{NoStop}%
\bibitem [{\citenamefont {Moshe}\ and\ \citenamefont
	{Zinn-Justin}(2003)}]{Moshe200369}%
\BibitemOpen
\bibfield  {author} {\bibinfo {author} {\bibfnamefont {M.}~\bibnamefont
		{Moshe}}\ and\ \bibinfo {author} {\bibfnamefont {J.}~\bibnamefont
		{Zinn-Justin}},\ }\href {\doibase
	http://dx.doi.org/10.1016/S0370-1573(03)00263-1} {\bibfield  {journal}
	{\bibinfo  {journal} {Physics Reports}\ }\textbf {\bibinfo {volume} {385}},\
	\bibinfo {pages} {69 } (\bibinfo {year} {2003})}\BibitemShut {NoStop}%
\bibitem [{\citenamefont {Cooper}\ \emph {et~al.}(2011)\citenamefont {Cooper},
	\citenamefont {Mihaila}, \citenamefont {Dawson}, \citenamefont {Chien},\ and\
	\citenamefont {Timmermans}}]{PhysRevA.83.053622}%
\BibitemOpen
\bibfield  {author} {\bibinfo {author} {\bibfnamefont {F.}~\bibnamefont
		{Cooper}}, \bibinfo {author} {\bibfnamefont {B.}~\bibnamefont {Mihaila}},
	\bibinfo {author} {\bibfnamefont {J.~F.}\ \bibnamefont {Dawson}}, \bibinfo
	{author} {\bibfnamefont {C.-C.}\ \bibnamefont {Chien}}, \ and\ \bibinfo
	{author} {\bibfnamefont {E.}~\bibnamefont {Timmermans}},\ }\href {\doibase
	10.1103/PhysRevA.83.053622} {\bibfield  {journal} {\bibinfo  {journal} {Phys.
			Rev. A}\ }\textbf {\bibinfo {volume} {83}},\ \bibinfo {pages} {053622}
	(\bibinfo {year} {2011})}\BibitemShut {NoStop}%
\bibitem [{\citenamefont {Steinhauer}\ \emph {et~al.}(2002)\citenamefont
	{Steinhauer}, \citenamefont {Ozeri}, \citenamefont {Katz},\ and\
	\citenamefont {Davidson}}]{PhysRevLett.88.120407}%
\BibitemOpen
\bibfield  {author} {\bibinfo {author} {\bibfnamefont {J.}~\bibnamefont
		{Steinhauer}}, \bibinfo {author} {\bibfnamefont {R.}~\bibnamefont {Ozeri}},
	\bibinfo {author} {\bibfnamefont {N.}~\bibnamefont {Katz}}, \ and\ \bibinfo
	{author} {\bibfnamefont {N.}~\bibnamefont {Davidson}},\ }\href {\doibase
	10.1103/PhysRevLett.88.120407} {\bibfield  {journal} {\bibinfo  {journal}
		{Phys. Rev. Lett.}\ }\textbf {\bibinfo {volume} {88}},\ \bibinfo {pages}
	{120407} (\bibinfo {year} {2002})}\BibitemShut {NoStop}%
\bibitem [{\citenamefont {Mihaila}\ \emph {et~al.}(2011)\citenamefont
	{Mihaila}, \citenamefont {Cooper}, \citenamefont {Dawson}, \citenamefont
	{Chien},\ and\ \citenamefont {Timmermans}}]{PhysRevA.84.023603}%
\BibitemOpen
\bibfield  {author} {\bibinfo {author} {\bibfnamefont {B.}~\bibnamefont
		{Mihaila}}, \bibinfo {author} {\bibfnamefont {F.}~\bibnamefont {Cooper}},
	\bibinfo {author} {\bibfnamefont {J.~F.}\ \bibnamefont {Dawson}}, \bibinfo
	{author} {\bibfnamefont {C.-C.}\ \bibnamefont {Chien}}, \ and\ \bibinfo
	{author} {\bibfnamefont {E.}~\bibnamefont {Timmermans}},\ }\href {\doibase
	10.1103/PhysRevA.84.023603} {\bibfield  {journal} {\bibinfo  {journal} {Phys.
			Rev. A}\ }\textbf {\bibinfo {volume} {84}},\ \bibinfo {pages} {023603}
	(\bibinfo {year} {2011})}\BibitemShut {NoStop}%
\bibitem [{\citenamefont {Davis}\ and\ \citenamefont
	{Blakie}(2006)}]{PhysRevLett.96.060404}%
\BibitemOpen
\bibfield  {author} {\bibinfo {author} {\bibfnamefont {M.~J.}\ \bibnamefont
		{Davis}}\ and\ \bibinfo {author} {\bibfnamefont {P.~B.}\ \bibnamefont
		{Blakie}},\ }\href {\doibase 10.1103/PhysRevLett.96.060404} {\bibfield
	{journal} {\bibinfo  {journal} {Phys. Rev. Lett.}\ }\textbf {\bibinfo
		{volume} {96}},\ \bibinfo {pages} {060404} (\bibinfo {year}
	{2006})}\BibitemShut {NoStop}%
\bibitem [{\citenamefont {Nair}(2006)}]{nair2006quantum}%
\BibitemOpen
\bibfield  {author} {\bibinfo {author} {\bibfnamefont {V.}~\bibnamefont
		{Nair}},\ }\href {https://books.google.com/books?id=rcr0fxDf8jsC} {\emph
	{\bibinfo {title} {Quantum Field Theory: A Modern Perspective}}},\ Graduate
Texts in Contemporary Physics\ (\bibinfo  {publisher} {Springer New York},\
\bibinfo {year} {2006})\BibitemShut {NoStop}%
\bibitem [{\citenamefont {Zee}(2010)}]{zee2010quantum}%
\BibitemOpen
\bibfield  {author} {\bibinfo {author} {\bibfnamefont {A.}~\bibnamefont
		{Zee}},\ }\href {https://books.google.com/books?id=n8Mmbjtco78C} {\emph
	{\bibinfo {title} {Quantum Field Theory in a Nutshell: (Second Edition)}}},\
In a Nutshell\ (\bibinfo  {publisher} {Princeton University Press},\ \bibinfo
{year} {2010})\BibitemShut {NoStop}%
\end{thebibliography}
%

\end{document}